\documentclass[journal,12pt,draftclsnofoot,onecolumn]{IEEEtran}

\makeatletter
\def\ps@headings{%
\def\@oddhead{\mbox{}\scriptsize\rightmark \hfil \thepage}%
\def\@evenhead{\scriptsize\thepage \hfil \leftmark\mbox{}}%
\def\@oddfoot{}
\def\@evenfoot{}}
\def\blfootnote{\xdef\@thefnmark{}\@footnotetext}
\makeatother
\usepackage{graphicx} 
\usepackage{epsfig} 
\usepackage{color}
\usepackage{mathptmx} 
\usepackage{times} 
\usepackage{latexsym}
\usepackage{amsmath} 
\usepackage{amssymb}  
\usepackage{subfigure}
\usepackage{algorithmic}
\usepackage{setspace}
\usepackage[bottom]{footmisc}
\newtheorem{theorem}{Theorem}
\newtheorem{conjecture}{Conjecture}
\newtheorem{lemma}{Lemma}
 
\newtheorem{definition}{Definition}
\newtheorem{corollary}{Corollary}

\newtheorem{remark}{Remark}
\title{\huge \bf
Optimal Computation of Symmetric Boolean Functions in Collocated Networks
}                                                                     
\author{\IEEEauthorblockN{Hemant Kowshik}\\
\IEEEauthorblockA{CSL and Department of ECE\\
University of Illinois Urbana-Champaign\\
Email: kowshik2@illinois.edu}\\
\and
\IEEEauthorblockN{P. R. Kumar}\\
\IEEEauthorblockA{CSL and Department of ECE\\
University of Illinois Urbana-Champaign\\
Email: prkumar@illinois.edu}
}
\begin{document}
\maketitle\blfootnote{This material is based upon work partially supported by AFOSR under Contract FA9550-09-0121, NSF under Contract No. CNS-1035378, Science \& Technology Center Grant CCF-0939370, Contract CNS-1035340, and Contract CNS-0905397, and USARO under Contract Nos. W911NF-08-1-0238 and W-911-NF-0710287. Any opinions, findings, and conclusions or recommendations expressed in this publication are those of the authors and do not necessarily reflect the views of the above agencies.}
\thispagestyle{empty}
\pagestyle{empty}
\begin{abstract}
\singlespacing
We consider collocated wireless sensor networks, where each node's transmissions can be heard by every other node. Each node has a Boolean measurement and the goal of the wireless sensor network is to compute a given Boolean function of these measurements. We first consider the worst case setting and study optimal block computation strategies for computing symmetric Boolean functions. We study three classes of functions: threshold functions, delta functions and interval functions. We provide exactly optimal strategies for the first two classes, and a scaling law order-optimal strategy with optimal preconstant for interval functions. We also extend the results to the case of integer measurements and certain integer-valued functions. We use lower bounds from communication complexity theory, and provide an achievable scheme using information theoretic tools. 

Next, we consider the case where nodes measurements are random and drawn from independent Bernoulli distributions. We address the problem of optimal function computation so as to minimize the expected total number of bits that are transmitted. In the case of computing a single instance of a Boolean threshold function, the problem reduces to one of determining the optimal order in which the nodes should transmit. We show the surprising result that the optimal order of transmissions depends in an extremely simple way on the values of previously transmitted bits, and the ordering of the marginal probabilities of the Boolean variables, according to the \emph{$k$-th least likely rule}: At any transmission, the node that transmits is the one that has the $k$-th least likely value of its Boolean variable, where $k$ reduces by one each time any node transmits a one. Initially the value of $k$ is ($n +1$ - Threshold). Surprisingly, the order of transmissions does not depend on the exact values of the probabilities of the Boolean variables, but only depends on their order.

The approach presented can be generalized to the case where each node has a block of measurements, though the resulting problem is somewhat harder, and we conjecture the optimal strategy. In the case of identically distributed measurements, we further show that the average-case complexity of block computation of a Boolean threshold function is $O(\theta)$, where $\theta$ is the threshold. We further show how to generalize to a pulse model of communication. One can also consider the related problem of approximate computation given a fixed number of bits. In this case, the optimal strategy is significantly different, and lacks an elegant characterization. However, for the special case of the parity function, we show that the greedy strategy is optimal.
\end{abstract}
\doublespacing
\section{INTRODUCTION}
Wireless sensor networks are composed of nodes with limited power and bandwidth, which can sense the environment and wirelessly communicate with each other to complete a collaborative task. However, in contrast to wireless \emph{data} networks, most wireless sensor network applications are not ininterested only in computing some relevant \textit{function} of the correlated data at distributed sensors. For instance, one might want to compute the mean temperature for environmental monitoring, or the maximum temperature in fire alarm systems. In order to extract the relevant information from enormous quantities of data generated by sensor nodes, one needs to design scalable algorithms and protocols. Thus, one needs to look beyond the data-forwarding paradigm and study how to design efficient in-network computation and communication strategies for functions of interest. 

The general problem of distributed function computation in wireless sensor networks presents several challenges. The broadcast nature of the wireless medium means that nodes have to deal with interference from other transmissions. This is typically resolved by a mechanism for distributed scheduling of transmissions. One of the consequences is that nodes which transmit later can exploit both previously received transmissions as well as the structure of the function to be computed, in order to create a more efficient description of their own data. Further, it is of interest to study the benefit of multi-round protocols, possibly involving complex interactions between nodes, versus single round protocols, where each node transmits only once. In the case where nodes have random data drawn from different distributions, there is the additional question regarding which node should transmit, since different nodes affect the function to different degrees. 

In this paper, we consider the collocated network scenario where all nodes can hear all transmissions. Its symmetry makes it a desirable starting point for studying random planar networks. At most one node is allowed to transmit at any time. Each node has a Boolean variable and we focus on the specific problem of symmetric Boolean function computation. We will focus on strategies which achieve function computation with zero error for \textit{all} sensor nodes. This is motivated by applications in distributed control and sensor-actuator networks, where each node needs to infer the state of the system in distributed fashion and respond in real time. We adopt a deterministic formulation of the problem of function computation, requiring zero error. We consider both the worst case setting and the average case setting where we impose a joint probability distribution on the node measurements. 

In Section \ref{sec_coll}, we consider the worst case setting, and address the problem of computing symmetric Boolean functions, which depend only on number of 1s, i.e., the ``type,". We study block computation strategies where nodes accummulate a block of measurements and employ block codes to achiever greater efficiency. The set of admissible strategies includes all interactive strategies, where a node may exchange several messages with other nodes. Since nodes can only transmit one at a time, the key challenge is for nodes to thoroughly exploit previous transmissions to compress their own data. We suppose that each node has a Boolean measurement and we wish to compute a given symmetric Boolean function of these measurements with zero error. We define three classes of functions, namely threshold functions which evaluate to $1$ if the number of $1$s exceeds a certain threshold, delta functions which evaluate to $1$ if the number of $1$s is exactly equal to a given value, and interval functions which evaluate to $1$ if the number of $1$s is between two given lower and upper values. For worst-case computation, we provide exactly optimal strategies for the first two classes, and a scaling law order-optimal strategy with optimal preconstant, as the number of nodes increases, for interval functions. Using these results, we can characterize the complexity of computing percentile type functions, which are of great interest. 

In our analysis, we use lower bounds from communication complexity theory, and provide an achievable scheme using information theoretic tools. Further, the approach presented can be generalized to compute functions of non-Boolean measurements, as shown in our treatment of general threshold functions and the $MAX$ function. While the proposed achievability strategy is exactly optimal for general threshold functions, it is only scaling law order-optimal for the $MAX$ function. 

In Section \ref{sec_coll_rand}, we address the case where the node measurements are independent and distributed according to given marginal Bernoulli distributions. Since the measurements are random, the evolution of the computation depends on the particular instances of measurements, and the time of termination of the computation is also accordingly random. We seek to minimize the total expected number of bits exchanged to achieve zero error computation. We primarily focus on optimal strategies for Boolean threshold functions. In the case of single instance computation, this is equivalent to determining the optimal ordering of nodes' transmissions so as to minimize the expected total number of bits exchanged. We present a surprising policy for ordering the transmissions and prove its optimality. The optimal policy is dynamic, depending in a particularly simple way on the previously transmitted bits, and on the relative ordering of the marginal probabilities, but, surprisingly, not on their values. The problem of optimally ordering transmissions of nodes is a sequential decision problem and can in principle be solved by dynamic programming. However, this would require solving the dynamic program for all thresholds and all probability distributions, which appears infeasible. We avoid this, and directly establish the optimal policy. The proposed solution permits a unified treatment of the problems of single instance computation, block computation and computation under alternate communication models.

In Section \ref{sec_two}, we turn our attention to the case where each node has a block of bits, and we seek to compute the Boolean threshold function for each instance of the block. This problem appears formidable due to the plethora of possibilities, and due to a far more complex class of interactive strategies for computation. However, for a certain natural restricted class of \emph{coherent} strategies, we can establish that an analogous policy is optimal, thus establishing an upper bound on the optimal cost. In order to establish a lower bound across all strategies, we propose the approach of calculating the minimum entropy over all valid protocol partitions which respect fooling set constraints. While this lower bound matches the upper bound for small examples, a proof has remained elusive. In Section \ref{sec_avg_thres}, we show that the average case complexity of computing a Boolean threshold function is $O(\theta)$, where $\theta$ is the threshold.

In Section \ref{sec_three}, we consider an alternate model of communication where nodes use pulses of unit energy to convey information. We generalize our proof technique and derive the optimal strategy for computing Boolean threshold functions under this model of communication. Finally, in Section \ref{sec_four}, we study the problem of approximate function computation given a fixed number of timeslots. We show that the optimal strategy for the approximate computation of threshold functions lacks the same elegant structure as present in the case of exact computation. However, for the special case of the parity function, we show that the logical strategy of first querying the node with maximum uncertainty, i.e., entropy, is optimal. 

\section{RELATED WORK}
The the problem of worst-case block function computation with zero error was formulated  in \cite{GiridharKumar}. 
The authors identified two classes of symmetric functions namely \textit{type-sensitive} functions exemplified by Mean, Median and Mode, and \textit{type-threshold} functions, exemplified by Maximum and Minimum. The maximum rates for computation of type-sensitive and type-threshold functions in random planar networks were shown to be $\Theta(\frac{1}{\log n})$ and $\Theta(\frac{1}{\log \log n})$ respectively, where $n$ is the number of nodes. If we impose a probability distribution on the node measurements, one can show that the average case complexity of computing type-threshold functions is $\Theta(1)$ \cite{KowshikKumar}.

In this paper, we address the problem where every node wishes to determine the value of the function. This approach naturally allows the use of tools from communication complexity \cite{KushiNisan}, where one seeks to find the minimum number of bits that must be exchanged in the worst case between two nodes, in order to achieve zero-error computation of a function of the node variables. The communication complexity of Boolean functions has been studied in \cite{Wegener}, \cite{OrlitskyElgamal}. In order to increase the computational efficiency, one can consider the \textit{direct-sum problem} \cite{KarchmerRazWigderson} where several instances of the problem are solved simultaneously. This block computation approach results in matching upper and lower bounds in the case of the Boolean AND function \cite{AhlswedeCai}. In this paper, we considerably generalize this result to derive the worst case complexity of computing Boolean threshold functions in collocated networks

If the measurements are drawn from some joint probability distribution and one is allowed block computation, we arrive at a distributed source coding problem with a fidelity criterion that is function-dependent, concerning which little is known. The problem of source coding with side information was studied in \cite{WynerZiv}. The extension of this approach to the problem of function computation with side information was studied in \cite{OrlitskyRoche}. The problem of interactive function computation in collocated networks has been studied in \cite{MaGuptaIshwar}.

Due to the broadcast nature of the wireless medium, two nodes which are close to each other cannot transmit simultaneously. Thus, nodes need to schedule their transmissions to avoid interfering with one another. The challenge now is to order nodes' transmissions so as to exploit the structure of the function, the side-information gained from previously transmitted bits, and the knowledge of the underlying distribution. Sequential decision making problems have been studied in various forms. The most well known problem of designing sequential experiments is the bandit problem \cite{Gittins}, \cite{GittinsJones}, \cite{Whittle}. One is given a bandit with multiple arms which offer different probabilistic payoffs. At each time-step, the player needs to choose which arm to play so as to maximize the expected long-term payoff. Thus, there is a tension between exploring which arms offer highest payoffs and exploiting them. Under the optimal strategy, each arm is assigned a dynamic allocation index and the arm with maximum index is chosen.

In \cite{ArrowPesotSobel}, an interesting problem in sequential decision making is studied, where, $n$ nodes have i.i.d. measurements, and a central agent wishes to know the identities of the nodes with the $k$ largest values. One is allowed questions of the type ``Is $X \geq t$'', to which the central agent receives the list of all nodes which satisfy the condition. Under this framework, the optimal recursive strategy of querying the nodes is found. A key difference in our formulation of the sequential decision making problem is that we are only allowed to query particular nodes, and not all nodes at once. The problem of minimizing the depth of decision trees for Boolean threshold queries is considered in \cite{AsherNewman}.
\section{Worst Case Computation of Symmetric Boolean Functions}\label{sec_coll}
In this section, we consider a collocated network with nodes $1$ through $n$, where each node's transmissions can be heard by every other node. Thus, the problem of medium access is resolved by allowing at most one node to transmit successfully at any time. Each node $i$ has a Boolean measurement $X_i \in \{0,1\}$, and \textit{every} node wants to compute the same function $f(X_1, X_2, \ldots, X_n)$ of the measurements. We seek to find communication schemes which achieve correct function computation at each node, with minimum worst-case total number of bits exchanged. We allow for the efficiencies of block computation, where each node $i$ has a block of $N$ independent measurements, denoted by $X_i^N$. Throughout this section, we suppose that collisions do not convey information thus restricting ourselves to \textit{collision-free strategies} as in \cite{GiridharKumar}. This means that for the $k^{th}$ bit $b_k$, the identity of the transmitting node $T_k$ depends only on previously broadcast bits $b_1, b_2, \ldots, b_{k-1}$, while the value of the bit it sends can depend arbitrarily on all previous broadcast bits as well as its block of measurements $X_{T_k}^{N}$.

It is important to note that all \textit{interactive} strategies are subsumed within the class of collision-free strategies. A collision-free strategy is said to achieve correct block computation if each node $i$ can  correctly determine the value of the function block $f^N(X_1, X_2, \ldots, X_n)$ using the sequence of bits $b_1, b_2, \ldots$ and its own measurement block $X_i^N$. Let $\mathcal{S}_N$ be the class of collision-free strategies for block length $N$ which achieve zero-error block computation, and let $C(f, S_N, N)$ be the worst-case total number of bits exchanged under strategy $S_N \in \mathcal{S}_N$. The worst-case per-instance complexity of computing a function $f(X_1, X_2, \ldots, X_n)$ is defined by
\begin{displaymath}
C(f) = \lim_{N \rightarrow \infty}\min_{S_N \in \mathcal{S}_N} \frac{C(f, S_N, N)}{N}.
\end{displaymath} 
We call this the \textit{broadcast computation complexity} of the function $f$.

We begin by recalling Theorem $1$ in \cite{AhlswedeCai}, which states that the complexity of computing the AND function of two variables is $\log_{2}3$ bits. In Section \ref{sec_and_function}, we generalize this result to obtain the broadcast communication complexity of the AND function of $n$ variables. In Section \ref{sec_threshold_function}, we derive the broadcast computation complexity for the more general class of \textit{threshold functions}. For this class of functions, we devise an achievable strategy which involves each node transmitting in turn, using a prefix-free codebook, and a lower bound based on fooling sets. It is interesting to note that the optimal strategy requires no back-and-forth interaction between nodes. In Sections \ref{sec_delta_function} and \ref{sec_interval_function}, we extend this approach to derive the broadcast computation complexity of computing \textit{delta functions} and \textit{interval functions} respectively. 

Finally, in Section \ref{sec_gen_alpha}, we present some extensions to the case of non-Boolean measurements and to the case of non-Boolean functions. Using the intuition gained from the Boolean case, we show how the achievability scheme and fooling set lower bounds can be adapted. In particular we study general threshold functions and the $MAX$ function. 

\subsection{Complexity of computing the AND function}\label{sec_and_function}
We consider now the specific problem of computing the AND function, which is $1$ if all its arguments are $1$, and $0$ otherwise. Consider a collocated network with $n$ nodes, each of which wants to compute the AND function of $n$ variables, denoted $\wedge(X_1, X_2, \ldots, X_n)$. For the case where $n=2$, we know from Theorem 1 in \cite{AhlswedeCai} that the broadcast communication complexity of computing the AND function is $\log_{2} 3$ bits. We have the following result for general $n$.
\begin{theorem}\label{thm_multiple_node_and}
For any strategy $S_N$, 
\begin{displaymath}
C(X_1 \wedge X_2 \ldots X_n, S_N, N) \geq N \log_{2}(n+1).
\end{displaymath}
Further, there exists a strategy $S_N^*$ such that 
\begin{displaymath}
C(X_1 \wedge X_2 \ldots X_n, S_N^*, N) \leq \lceil N \log_{2}(n+1)\rceil + (n-2).
\end{displaymath}
Thus, the complexity of the multiple node AND function is given by $C(\wedge(X_1, X_2, \ldots X_n)) = \log_{2}(n+1)$.
\end{theorem}
\textbf{Proof of Achievability:} The upper bound is established using induction on the number of nodes $n$. From Theorem 1 in \cite{AhlswedeCai}, the result is true for $n=2$ which is the basis step. Suppose the result is true for a collocated network of $(n-1)$ nodes. Consider an achievable scheme in which node $n$ transmits first, using a prefix free codebook. Let the length of the codeword transmitted be $l(X_n^N)$. After this transmission, the function is determined for the instances where $X_n = 0$. For the instances where $X_n = 1$, the remaining $(n-1)$ nodes need to compute $\wedge(X_1, X_2, \ldots, X_{n-1})$. From the induction hypothesis, we know that this can be done using $\lceil w(X_n^N)\log_{2}n\rceil + (n-3)$ bits. Thus the worst-case total number of bits exchanged is $L := \max_{X_n^N}(l(X_n^N) + \lceil w(X_n^N)\log_{2}n  + (n-3)\rceil)$. As before, we want to minimise this quantity subject to the Kraft inequality. Consider a prefix-free codebook for node $n$ which satisfies 
\begin{displaymath}
l(X_n^N) = \lceil N\log_{2}(n+1)\rceil + (n-2) - \lceil w(X_n^N)\log_{2}n\rceil - (n-3) \end{displaymath}
This satisfies Kraft inequality since
\begin{displaymath}
\sum_{X_n^N}2^{\lceil w(X_n^N)\log_{2}n\rceil} \leq \sum_{X_n^N}2^{w(X_n^N)\log_{2}n + 1} \leq 2(n+1)^N \leq 2^{\lceil N\log_{2}(n+1)\rceil + 1}
\end{displaymath}
\textbf{Proof of lower bound:} The lower bound is shown by constructing a \textit{fooling set} \cite{KushiNisan} of the appropriate size. We digress briefly to introduce the concept of fooling sets in the context of two-party communication complexity \cite{KushiNisan}. Consider two nodes $X$ and $Y$, each of which take values in finite sets $\mathcal{X}$ and $\mathcal{Y}$, and both nodes want to compute some function $f(X,Y)$ with zero error. 
\begin{definition}[Fooling Set]
A set $E \subseteq \mathcal{X} \times \mathcal{Y}$ is said to be a fooling set, if for any two distinct elements $(x_1, y_1), (x_2, y_2)$ in $E$, we have either 
\begin{itemize}
\item $f(x_1, y_1) \neq f(x_2, y_2)$, or
\item $f(x_1, y_1) = f(x_2, y_2)$, but either $f(x_1,y_2) \neq f(x_1, y_1)$ or $f(x_2,y_1) \neq f(x_1,y_1)$.
\end{itemize}
\end{definition}
Given a fooling set $E$ for a function $f(X_1, X_2)$, we have $C(f(X_1, X_2)) \geq \log_{2}|E|$. We have described two dimensional fooling sets above. The extension to multi-dimensional fooling sets is straightforward and gives a lower bound on the communication complexity of the function $f(X_1, X_2, \ldots, X_n)$. 

We need to devise a subset of the set of all $n \times N$ measurement matrices which is a valid fooling set. Consider the subset $E$ of measurement matrices which are only comprised of columns which sum to $(n-1)$ or $n$. Since there are $N$ columns, there are $(n+1)^N$ such matrices. Let $M_1$, $M_2$ be two distinct matrices in this subset. If $f^{N}(M_1) \neq f^{N}(M_2)$, then we are done. Suppose not. Then there must exist one instance where the function evaluates to zero and for which $M_1$ and $M_2$ have different columns. Let us suppose $M_1$ has $1_n - e_i$ and $M_2$ has $1_n - e_j$. Now if we replace the $i^th$ row of $M_1$ with the $i^th$ row of $M_2$, the resulting measurement matrix, say $M^{*}$ is such that $f(M^{*}) \neq f(M_1)$. Thus, the set $E$ is a valid fooling set. From the fooling set lower bound, we have, for \textit{any} strategy $S_N \in \mathcal{S}_N$, we must have $C(\wedge(X_1, X_2), S_N, N) \geq N\log_{2}3$ implying that $C(f) \geq \log_{2}3$. This concludes the proof of Theorem \ref{thm_multiple_node_and}. $\Box$

By symmetry, we can derive the complexity of the OR function, which is defined to be $0$ if all its arguments are $0$, and $1$ otherwise. Consider a collocated network with $n$ nodes, each of which wants to compute the OR function, denoted by $\vee(X_1, X_2, \ldots, X_n)$.
\begin{corollary}\label{cor_multiple_node_or}
The complexity of the OR function is given by $C(\vee(X_1, X_2, \ldots, X_n)) = \log_{2}(n+1)$, since we can view it as $\overline{\wedge(\overline{X}_1 , \overline{X}_2, \ldots, \overline{X}_n)}$, by deMorgan's laws.
\end{corollary}
\textbf{Note:} Throughout the rest of this section, for ease of exposition, we will ignore the fact that terms like $N\log_{2}(n+1)$ may not be integer. Since our achievability strategy involves each node transmitting exactly once, this will result in a maximum of one extra bit per node, and since we are amortizing this over a long block length $N$, it will not affect any of the results.
\subsection{Complexity of computing Boolean threshold functions} \label{sec_threshold_function}
\begin{definition}[Boolean threshold functions]
A Boolean threshold function $\Pi_{\theta}(X_1, X_2, \ldots, X_n)$ is defined as
\begin{displaymath}
\Pi_{\theta}(X_1, X_2, \ldots, X_n) = \left\{ \begin{array}{l} 1 \quad \textrm{if } \sum_{i}X_i \geq \theta \\ 0 \quad \textrm{otherwise.}\end{array}\right.
\end{displaymath}
\end{definition}
\begin{theorem}\label{thm_bool_threshold}
The complexity of computing a Boolean threshold function is $C(\Pi_{\theta}(X_1, X_2, \ldots X_n)) = \log_{2}\left(\begin{array}{c}n+1 \\ \theta \end{array}\right)$.
\end{theorem}
\textbf{Proof of Achievability:} The upper bound is established by induction on $n$. From Theorem \ref{thm_multiple_node_and} and Corollary \ref{cor_multiple_node_or}, the result is true for $n=2$ and for \textit{all} $1 \leq \theta \leq n$, which is the basis step. Suppose the upper bound is true for a collocated network of $(n-1)$ nodes, for all $1 \leq \theta \leq (n-1)$. Given a function $\Pi_{\theta}(X_1, X_2, \ldots, X_n)$ of $n$ variables, consider an achievable strategy in which node $n$ transmits first, using a prefix free codeword of length $l(X_n^N)$. After this transmission, nodes $1$ through $n-1$ can decode the block $X_n^N$. For the instances where $X_n = 0$, these $(n-1)$ nodes now need to compute $\Pi_{\theta}(X_1, X_2, \ldots, X_{n-1})$. For the instances where $X_n = 1$, the remaining $(n-1)$ nodes need to compute $\Pi_{\theta -1}(X_1, X_2, \ldots, X_{n-1})$. From the induction hypothesis, we have optimal strategies for computing these functions. Let $w^{i}(X_n^N)$ denote the number of instances of $i$ in the block $X_n^N$. Under the above strategy, the worst-case total number of bits exchanged is
\begin{displaymath}
L = \max_{X_n^N} \left(l(X_n^N) + w^{0}(X_n^N) \log_{2}\left(\begin{array}{c} n \\ \theta \end{array}\right) + w^{1}(X_n^N) \log_{2}\left(\begin{array}{c} n \\ \theta -1 \end{array}\right) \right). 
\end{displaymath}
We want to minimise this quantity subject to the Kraft inequality. Consider a prefix-free codebook which satisfies 
\begin{displaymath}
l(X_n^N) = N\log_{2}\left(\begin{array}{c} n+1 \\ \theta \end{array}\right) - w^{0}(X_n^N) \log_{2}\left(\begin{array}{c} n \\ \theta \end{array}\right) - w^{1}(X_n^N) \log_{2}\left(\begin{array}{c} n \\ \theta -1 \end{array}\right).
\end{displaymath}
This assignment of codelengths satisfies the Kraft inequality since
\begin{eqnarray}
\sum_{X_n^N}2^{-l(X_n^N)} & = & \left(\begin{array}{c}n+1 \\ \theta \end{array}\right)^{-N}\sum_{X_n^N}\left(\begin{array}{c}n \\ \theta \end{array}\right)^{w^{0}(X_n^N)} \left(\begin{array}{c}n \\ \theta -1 \end{array}\right)^{w^{1}(X_n^N)} \nonumber \\
& = & \left(\begin{array}{c}n+1 \\ \theta \end{array}\right)^{-N}\left[\left(\begin{array}{c}n \\ \theta \end{array}\right) + \left(\begin{array}{c}n \\ \theta -1 \end{array}\right)\right]^{N} = 1. \nonumber
\end{eqnarray}
Hence there exists a prefix-free code which satisfies the specified codelengths, and we have $L =   N \log_{2}\left(\begin{array}{c}n+1 \\ \theta \end{array}\right)$, which proves the induction step.\\
\textbf{Proof of lower bound:} We need to find a subset of the set of all $n \times N$ measurement matrices which is a valid fooling set. Consider the subset $E$ of measurement matrices which consist of only columns which sum to $(\theta-1)$ or $\theta$. Since there are $N$ columns, there are $\left[\left(\begin{array}{c}n \\ \theta \end{array}\right) + \left(\begin{array}{c}n \\ \theta -1 \end{array}\right)\right]^{N}$ such matrices. We claim that the set $E$ is a valid fooling set. Let $M_1$, $M_2$ be two distinct matrices in this subset. If $f^{N}(M_1) \neq f^{N}(M_2)$, then we are done. Suppose not. Then there must exist at least one column at which $M_1$ and $M_2$ disagree, say $M_1^{(j)} \neq M_2^{(j)}$. However, both $M_1^{(j)}$ and $M_2^{(j)}$ have the same number of ones. Thus there must exist some row, say $i^*$, where $M_1^{(j)}$ has a zero, but $M_2^{(j)}$ has a one.
\begin{itemize}
\item[(i)] Suppose $f(M_1^{(j)}) = f(M_2^{(j)}) = 0$. Then, consider the matrix $M_1^*$ obtained by replacing the $i^*$th row of $M_1$ with the $i^*$th row of $M_2$. The $j^{th}$ column of $M_1^*$ has $\theta$ ones, and hence $f(M_1^{*(j)}) = 1$. Hence we have $f(M_1^*) \neq f(M_1)$.
\item[(ii)] Suppose $f(M_1^{(j)}) = f(M_2^{(j)}) = 1$. Then, consider the matrix $M_2^*$ obtained by replacing the $i^*$th row of $M_2$ with the $i^*$th row of $M_1$. The $j^{th}$ column of $M_2^*$ has $\theta -1$ ones, and hence $f(M_2^{*(j)}) = 1$. Hence we have $f(M_2^*) \neq f(M_2)$.
\end{itemize}

Thus, the set $E$ is a valid fooling set. From the fooling set lower bound, for \textit{any} strategy $S_N \in \mathcal{S}_N$, we must have $C(\Pi_{\theta}(X_1, X_2, \ldots, X_n), S_N, N) \geq N\log_{2}\left(\begin{array}{c}n+1 \\ \theta \end{array}\right)$ implying that $C(\Pi_{\theta}(X_1, X_2, \ldots, X_n)) \geq \log_{2}\left(\begin{array}{c}n+1 \\ \theta \end{array}\right)$. $\Box$ 
\subsection{Complexity of Boolean delta functions}\label{sec_delta_function}
\begin{definition}[Boolean delta function]
A Boolean delta function $\Pi_{\{\theta\}}(X_1, X_2, \ldots, X_n)$ is defined as:
\begin{displaymath}
\Pi_{\{\theta\}}(X_1, X_2, \ldots, X_n) = \left\{\begin{array}{l} 1 \quad \textrm{if } \sum_{i}X_i = \theta \\ 0 \quad \textrm{otherwise.}\end{array}\right.
\end{displaymath}
\end{definition}
\begin{theorem}\label{thm_bool_delta}
The complexity of computing $\Pi_{\{\theta\}}(X_1, X_2, \ldots, X_n)$ is given by
\begin{displaymath} 
C(\Pi_{\{\theta\}}(X_1, X_2, \ldots, X_n)) = \log_{2}\left[\left(\begin{array}{c}n+1 \\ \theta \end{array}\right) + \left(\begin{array}{c}n \\ \theta + 1 \end{array}\right) \right]. 
\end{displaymath}
\end{theorem}
\textbf{Sketch of Proof:} The proof of achievability follows from an inductive argument as before. The fooling set $E$ consists of measurement matrices composed of only columns which sum up to $\theta -1$, $\theta$ or $\theta + 1$. Thus the size of the fooling set is \begin{displaymath}
\left[\left(\begin{array}{c}n \\ \theta -1 \end{array}\right) + \left(\begin{array}{c}n \\ \theta \end{array}\right) + \left(\begin{array}{c}n \\ \theta + 1 \end{array}\right) \right]^{N}. \Box
\end{displaymath}
\subsection{Complexity of computing Boolean interval functions}\label{sec_interval_function}
A Boolean \textit{interval function} $\Pi_{[a,b]}(X_1, \ldots, X_n)$ is defined as:
\begin{displaymath}
\Pi_{[a,b]}(X_1, X_2, \ldots, X_n) = \left\{\begin{array}{l} 1 \quad \textrm{if } a \leq \sum_{i}X_i \leq b \\ 0 \quad \textrm{otherwise.}\end{array} \right.
\end{displaymath}
A naive strategy to compute the function $\Pi_{[a,b]}(X_1, \ldots, X_n)$ is to compute the threshold functions $\Pi_{a}(X_1, \ldots, X_n)$ and $\Pi_{b+1}(X_1, X_2, \ldots, X_n)$. However, this strategy gives us more information than we seek, i.e., if $\sum_{i} X_i \in [a,b]^{C}$, then we also know if $\sum_{i} X_i < a$, which is superfluous information and perhaps costly to obtain. Alternately, we can derive a strategy which explicitly deals with intervals, as against thresholds. This strategy has significantly lower complexity.
\begin{theorem}\label{thm_bool_interval}
The complexity of computing a Boolean interval function $\Pi_{[a,b]}(X_1, X_2, \ldots, X_n)$ with $a + b \leq n$ is bounded as follows:
\begin{multline}
\log_{2}\left[\left(\begin{array}{c}n+1 \\ b+1 \end{array}\right) + \left(\begin{array}{c}n \\ a-1 \end{array}\right)\right] \leq C(\Pi_{[a,b]}(X_1, X_2, \ldots X_n)) \\
\leq \log_{2}\left[ \left(\begin{array}{c}n+1 \\ b+1 \end{array}\right) + (b-a + 1)\left(\begin{array}{c}n \\ a-1 \end{array}\right)\right]. \label{int_bound_caseone} 
\end{multline}
The complexity of computing a Boolean interval function $\Pi_{[a,b]}(X_1, \ldots, X_n)$ with $a +b \geq n$ is bounded as follows:
\begin{multline}
\log_{2}\left[\left(\begin{array}{c}n+1 \\ a \end{array}\right) + \left(\begin{array}{c}n \\ b+1 \end{array}\right)\right] \leq C(\Pi_{[a,b]}(X_1, X_2, \ldots X_n)) \\
\leq \log_{2}\left[ \left(\begin{array}{c}n+1 \\ a \end{array}\right) + (b-a+1)\left(\begin{array}{c}n\\ b+1 \end{array}\right)\right]. \label{int_bound_casetwo} 
\end{multline}
\end{theorem}
\textbf{Proof of lower bound:} Suppose $a+b \leq n$. Consider the subset $E$ of measurement matrices which consist of only columns which sum to $(a-1)$, $b$ or $(b+1)$. We claim that the set $E$ is a valid fooling set. Let $M_1$, $M_2$ be two distinct matrices in this subset. If $f^{N}(M_1) \neq f^{N}(M_2)$, we are done. Suppose not. Then there must exist at least one column at which $M_1$ and $M_2$ disagree, say $M_1^{(j)} \neq M_2^{(j)}$. 
\begin{itemize}
\item[(i)] Suppose $f(M_1^{(j)}) = f(M_2^{(j)}) = 1$. Then, both $M_1^{(j)}$ and $M_2^{(j)}$ have exactly $b$ $1$s. Thus there exists some row, say $i^*$, where $M_1^{(j)}$ has a $0$, but $M_2^{(j)}$ has a $1$. Consider the matrix $M_1^*$ obtained by replacing the $i^*$th row of $M_1$ with the $i^*$th row of $M_2$. The $j^{th}$ column of $M_1^*$ has $(b+1)$ $1$s, and hence $f(M_1^{*(j)}) = 0$, which means $f(M_1^*) \neq f(M_1)$. 
\item[(ii)] Suppose $f(M_1^{(j)}) = f(M_2^{(j)}) = 0$. If both $M_1^{(j)}$ and $M_2^{(j)}$ have the same number of $1$s, then the same argument as in (i) applies. However, if $M_1^{(j)}$ has $(a-1)$ $1$s and $M_2^{(j)}$ has $(b+1)$ $1$s, then there exists some row $i^*$ where $M_1^{(j)}$ has a $0$, but $M_2^{(j)}$ has a $1$. Then, the matrix $M_2^*$ obtained by replacing the $i^*$th row of $M_2$ with the $i^*$th row of $M_1$ is such that $f(M_2^*) \neq f(M_2)$. 
\end{itemize}

Thus, the set $E$ is a valid fooling set and $|E| = \left[\left(\begin{array}{c}n \\ b+1 \end{array}\right) + \left(\begin{array}{c}n \\ a -1 \end{array}\right) + \left(\begin{array}{c}n \\ b \end{array}\right)\right]^{N}$. This gives us the required lower bound in (\ref{int_bound_caseone}).

For the case where $a +b \geq n$, we consider the fooling set $E'$ of matrices which are comprised of only columns which sum to $a-1$, $a$ or $b+1$. This gives us the lower bound in (\ref{int_bound_casetwo}).\\
\textbf{Proof of achievability:} Consider the general strategy for achievability where node $n$ transmits a prefix-free codeword of length $l(X_1^N)$, leaving the remaining $(n-1)$ nodes the task of computing a residual function. This approach yields a recursion for computing the complexity of interval functions.
\begin{displaymath}
C(\Pi_{[a,b]}(X_1, \ldots, X_n)) \leq \log_{2}\left[2^{C(\Pi_{[a-1, b-1]}(X_1, \ldots, X_{n-1}))} + 2^{C(\Pi_{[a,b]}(X_1, \ldots ,X_{n-1})}\right].
\end{displaymath}

The boundary conditions for this recursion are obtained from the result for Boolean threshold functions in Theorem \ref{thm_bool_threshold}. We could simply solve this recursion computationally, but we want to study the behaviour of the complexity as we vary $a$, $b$ and $n$. Define $h_{a,b,n} := 2^{C(\Pi_{[a,b]}(X_1, \ldots, X_n))}$. We have the following recursion for $h(a,b,n)$
\begin{equation}\label{f_recursion}
h(a,b,n) \leq h(a-1,b-1,n-1) + h(a,b,n-1). 
\end{equation}
We proceed by induction on $n$. From Theorems \ref{thm_bool_threshold} and \ref{thm_bool_delta}, the upper bounds in (\ref{int_bound_caseone}) and (\ref{int_bound_casetwo}) are true for $n=2$ and all intervals $[a,b]$. Suppose the upper bound is true for all intervals $[a,b]$ for $(n-1)$ nodes. Consider the following cases.
\begin{itemize}
\item[(i)] Suppose $a + b \leq n-1$. Substituting the induction hypothesis in (\ref{f_recursion}), we get
\begin{eqnarray}
h(a,b,n) & \leq & \left(\begin{array}{c}n \\b \end{array}\right) + (b-a +1)\left(\begin{array}{c}n-1 \\a-2 \end{array}\right) \nonumber \\
& & + \left(\begin{array}{c}n \\b+1 \end{array}\right) + (b-a+1)\left(\begin{array}{c}n-1 \\a-1 \end{array}\right) \nonumber \\
& =  & \left(\begin{array}{c}n+1 \\b+1 \end{array}\right) + (b-a+1)\left(\begin{array}{c}n \\a-1 \end{array}\right). \nonumber
\end{eqnarray}
\item[(ii)] Suppose $a+b \geq n+1$. Proof is similar to case (i).
\item[(iii)] Suppose $a + b = n$. Substituting the induction hypothesis in (\ref{f_recursion}), we get
\begin{eqnarray}
h(a,b,n) & \leq & \left(\begin{array}{c}n \\b \end{array}\right) + (b-a+1)\left(\begin{array}{c}n-1 \\a-2 \end{array}\right) \nonumber \\ 
& & + \left(\begin{array}{c}n \\a \end{array}\right) + (b-a+1)\left(\begin{array}{c}n-1 \\b+1 \end{array}\right) \nonumber \\
& \leq & \left(\begin{array}{c}n+1 \\a \end{array}\right) + (b-a+1)\left(\begin{array}{c}n \\ b+1 \end{array}\right). \nonumber 
\end{eqnarray}
where some steps have been omitted in the proof of the last inequality. This establishes the induction step and completes the proof. $\Box$
\end{itemize}
\subsubsection{Discussion of Theorem \ref{thm_bool_interval}}
\begin{itemize}
\item[(a)] The gap between the lower and upper bounds in (\ref{int_bound_caseone}) and (\ref{int_bound_casetwo}) is \textit{additive}, and is upper bounded by $\log_{2}(b-a + 2)$ which is $\log_{2}(n+2)$ in the worst case.
\item[(b)] For fixed $a$ and $b$, as the number of nodes increases, we have $a+b \leq n$ for large enough $n$.  Consider the residual term, $(b-a+1)\left(\begin{array}{c}n \\a-1 \end{array}\right)$ on the RHS in (\ref{int_bound_caseone}). We have
\begin{displaymath}
(b-a+1)\left(\begin{array}{c}n \\a-1 \end{array}\right) = o\left(\left(\begin{array}{c}n+1 \\ b+1 \end{array}\right)\right).
\end{displaymath}
Hence, $C(\Pi_{[a,b]}(X_1, \ldots, X_n)) = \log_{2}\left(\left(\begin{array}{c}n+1 \\ b+1 \end{array}\right)(1 + o(1))\right)$. Thus, for any fixed interval $[a,b]$, we have derived an order optimal strategy with optimal preconstant. The orderwise complexity of this strategy is the same as that of the threshold function $\Pi_{b+1}(X_1, \ldots, X_n)$.  Similarly, we can derive order optimal strategies for computing $C(\Pi_{[n-a,n-b]}(X_1, \ldots, X_n))$ and $C(\Pi_{[a,n-b]}(X_1, \ldots, X_n))$, for fixed $a$ and $b$. 
\item[(c)] Consider a \textit{percentile} type function where $[a,b] = [\alpha n, \beta n]$, with $(\alpha + \beta) \leq 1$. Using Stirling's approximation, we can still show that
\begin{displaymath}
(\beta - \alpha)n\left(\begin{array}{c}n \\ \alpha n -1 \end{array}\right) = o\left(\left(\begin{array}{c}n+1 \\ \beta n+1 \end{array}\right)\right).
\end{displaymath}
Thus we have derived an order optimal strategy with optimal preconstant for percentile functions.
\item[(d)] Consider the function $f := \Pi_{\cup_{i} [a_i, b_i]}(X_1, \ldots, X_n)$ where the intervals $[a_i, b_i]$ are disjoint, and may be fixed or percentile type. We can piece together the result for single intervals and show that 
\begin{displaymath}
C(f(X_1, \ldots, X_n)) = \log_{2}\left(\sum_{i=1}^{m}g(a_i,b_i,n)(1 + o(1))\right).
\end{displaymath}
\begin{displaymath}
\textrm{where } g(a_i,b_i,n) = \left\{\begin{array}{ll} \left(\begin{array}{c}n+1 \\ b_i+1 \end{array}\right) \textrm{ if } a_i + b_i \leq n\\
\left(\begin{array}{c}n+1 \\ a_i \end{array}\right) \textrm{ if } a_i + b_i \geq n. \end{array} \right.
\end{displaymath}
\end{itemize}
\subsection{Extension to general alphabets}\label{sec_gen_alpha}
In Sections \ref{sec_threshold_function} - \ref{sec_interval_function}, we have studied optimal strategies for computing threshold functions, delta functions and interval functions of Boolean measurements. In this section, we will show that these results can be generalized to the case where nodes have general integer alphabets, i.e., $X_i \in \{0, 1, \ldots, m_i\}$. The proofs are lengthier in this case, and to maintain clarity of presentation, we will focus on threshold functions and the $MAX$ function. 
\subsubsection{Complexity of General Threshold Functions}
Consider a collocated network of $n$ nodes, where node $i$ has measurement $X_i \in \{0, 1, \ldots, m_i\}$.
\begin{definition}\label{General Threshold Function}
A general threshold function $\Pi_{\theta}(X_1, X_2, \ldots, X_n)$ is defined as below.
\begin{displaymath}
\Pi_{\theta}(X_1, X_2, \ldots, X_n) := \left\{ \begin{array}{ll} 1 \quad \textrm{if } \sum_{i=1}^{n}X_i \geq \theta \\ 0 \quad \textrm{otherwise}\end{array}\right. .
\end{displaymath}
\end{definition}
We employ the same notation as for Boolean threshold functions, which constitute a special case of general threshold functions.
\begin{theorem}\label{thm_gen_threshold}
The complexity of computing $\Pi_{\theta}(X_1, \ldots, X_n)$ is given by
\begin{eqnarray} 
C(\Pi_{\theta}(X_1, \ldots, X_n)) & = & \log_{2}\left( \left[Y^{\theta}\right] + \left[Y^{\theta -1 }\right] \left(\prod_{i = 1}^{n}\frac{1 - Y^{m_i + 1}}{1-Y} \right)\right). \nonumber \\
& = & \log_{2}\left( \left[Y^{\theta}\right] + \left[Y^{\theta -1 }\right] \left(\prod_{i = 1}^{n}(1 - Y^{m_i + 1}) \right)\left(\sum_{k = 1}^{\infty}\left(\begin{array}{cc} n+k-1\\  n-1\end{array}\right)Y^k \right)\right) \nonumber
\end{eqnarray}
where the notation $[Y^{\theta}]$ refers to the coefficient of $Y^{\theta}$ in the expression on the RHS.
\end{theorem}
\textbf{Proof: } The proof proceeds by induction on the number of nodes $n$. From Theorem $1$ in \cite{KowshikKumar_Tree}, we know that the result is true for $n =2$ and all choices of $m_1, m_2$ and $\theta$. This serves as a basis step for the induction. Let us suppose the result is true for a collocated network of $n-1$ nodes and all choices of $m_1, m_2, \ldots, m_{n-1}$ and $\theta$. We now proceed to prove the result for a network of $n$ nodes. 

We specify a strategy $S_N^*$ in which node $n$ transmits first. As described in \cite{KowshikKumar_Tree}, the optimal strategy consists of two stages, namely separation and coding. We begin by identifying the symbols in  $\{0, 1, \ldots, m_{n}\}$ that need to be \textit{separated} by node $n$. Let $\tilde{X}_n$ be the mapping of $X_n$ to the reduced alphabet given by $\{a_n,\ldots,b_n\}$. Subsequently, we construct a prefix-free codeword on the reduced alphabet.  Let the length of the codeword transmitted be $l(X_1^N)$. At the end of this transmission, the remaining $n-1$ nodes need to compute a residual threshold function for each instance of the block. For example, if $X_n = k$, we are left with the task of computing $\Pi_{\theta -k}(X_1, \ldots, X_{n-1})$. By the induction hypothesis, there is an achievable strategy to compute this residual threshold function, with complexity $C(\Pi_{\theta -k}(X_1, \ldots, X_{n-1}))$. Thus the worst case total number of bits exchanged under this strategy is given by
\begin{multline*}
L := \max_{\tilde{X}_n^N} (l(\tilde{X}_n^N) + w^{a_n}(\tilde{X}_n^N)C(\Pi_{\theta - a_n}(X_1, \ldots, X_{n-1})) + w^{a_n + 1}(\tilde{X}_n^N)C(\Pi_{\theta - a_n -1}(X_1, \ldots, X_{n-1}))  \\
+ \ldots + w^{b_n}(\tilde{X}_n^N)C(\Pi_{\theta - b_n}(X_1, \ldots, X_{n-1})),
\end{multline*}
where $w^{j}(\tilde{X}_n^N)$ is the number of instances in the block where $\tilde{X}_n = j$. Our objective is to find the smallest $L$ that satisfies the Kraft inequality for prefix free codes, which states that $\sum_{\tilde{X}_n^N}2 ^ {-l(\tilde{X}_n^N)} \leq 1$. From the definition of $L$, we can lower bound the LHS of the Kraft inequality.
\begin{displaymath}
\sum_{X_n^N}2 ^ {-l(\tilde{X}_n^N)} \geq 2^{L}\sum_{\tilde{X}_n^N}\left(2^{-w^{a_n}(\tilde{X}_n^N)C(\Pi_{\theta - a_n}(X_1, \ldots, X_{n-1}))}\ldots 2^{-w^{b_n}(\tilde{X}_n^N)C(\Pi_{\theta-b_n}(X_1, \ldots, X_{n-1}))} \right).
\end{displaymath}
From the induction hypothesis, we have that 
\begin{displaymath}
C(\Pi_{\theta - k}(X_1, \ldots, X_{n-1})) =\log_ {2}\left(\left[Y^{\theta - k}\right] + \left[Y^{\theta - k - 1}\right] \left(\prod_{i = 1}^{n-1}\frac{(1 - Y^{m_i + 1})}{1-Y} \right)\right)
\end{displaymath}
Thus, the smallest feasible value of $L$ is given by
\begin{eqnarray}
2^L & = & \sum_{\tilde{X}_n^N} \left(\left[Y^{\theta -a_n }\right] + \left[Y^{\theta - a_n -1}\right] \prod_{i=1}^{n-1}\left(\frac{1 - Y^{m_i + 1}}{1-Y} \right)\right)^{w^{a_n}(\tilde{X}_n^N)} \cdot \ldots 
\nonumber \\
& &\qquad \qquad \qquad \qquad \qquad \qquad \cdot \ldots \cdot \left(\left[Y^{\theta - b_n }\right] + \left[Y^{\theta - b_n - 1}\right]  \prod_{i=1}^{n-1}\left(\frac{1 - Y^{m_i + 1}}{1-Y} \right)\right)^{w^{b_n}(\tilde{X}_n^N)} \nonumber \\
& = & \left(\sum_{k = a_n}^{b_n}\left(\left[Y^{\theta - k }\right] + \left[Y^{\theta - k - 1}\right]  \prod_{i=1}^{n-1}\left(\frac{1 - Y^{m_i + 1}}{1-Y} \right)\right)\right)^{N} \nonumber \\
& = &  \left(\sum_{k = 0}^{m_n}\left(\left[Y^{\theta - k }\right] + \left[Y^{\theta - k - 1}\right]  \prod_{i=1}^{n-1}\left(\frac{1 - Y^{m_i + 1}}{1-Y} \right)\right)\right)^{N} \label{adjust_one} \\
& = & \left(\left[Y^{\theta} \right] +  \left[Y^{\theta - 1}\right](1 + Y + \ldots + Y^{m_n}) \prod_{i=1}^{n-1}\left(\frac{1 - Y^{m_i + 1}}{1-Y} \right)\right)^{N} \nonumber \\
& = & \left(\left[Y^{\theta} \right] +  \left[Y^{\theta - 1}\right]  \prod_{i=1}^{n}\left(\frac{1 - Y^{m_i + 1}}{1-Y} \right)\right)^{N}. \nonumber \\
L & = & N \log_{2}\left(\left[Y^{\theta} \right] +  \left[Y^{\theta - 1}\right]  \prod_{i=1}^{n}\left(\frac{1 - Y^{m_i + 1}}{1-Y} \right)\right).
\end{eqnarray}
where (\ref{adjust_one}) follows from the fact that for $k < a_n$ and $k  > b_n$, the coefficients of $Y^{\theta - k }$ and $Y^{\theta - k - 1}$ are both zero. Thus, we have derived an upper bound on the complexity of computing general threshold functions in collocated networks. \\
\textbf{Proof of lower bound:}  We need to find a subset of the set of all $n \times N$ measurement matrices which is a valid fooling set. Consider the subset $E$ of measurement matrices which are made up only of the column vectors which sum to $(\theta-1)$ or $\theta$. Consider two distinct measurement matrices $M_1, M_2 \in E$. Let $f^N(M_1)$ and $f^N(M_2)$ be the block function values obtained from these two matrices. If $f^N(M_1) \neq f^N(M_2)$, we are done. Let us suppose $f^N(M_1) = f^N(M_2)$, and note that since $M_1 \neq M_2$, there must exist one column, say column $j$, where $M_1$ and $M_2$ differ. However, since $f^N(M_1) = f^N(M_2)$, each column of $M_1$ must sum to the same value as the corresponding column in $M_2$. Thus there must exist rows $i_1$ and $i_2$ such that $M_1(i_1, j) < M_2(i_1, j)$ and $M_1(i_2, j) < M_2(i_2, j)$. 
\begin{itemize}
\item If column $j$ in $M_1$ and $M_2$ sum to $\theta -1$, then consider the new measurement matrix $M^{*}$ got by replacing the $i_1^{th}$ row of $M_1$ with the $i_1^{th}$ row of $M_2$. The $j^{th}$ column of $M^{*}$ sums to a value that is greater than $\theta -1$. Thus, we have $f(M^{*}) \neq f(M_1)$.
\item If column $j$ in $M_1$ and $M_2$ sum to $\theta$, then consider the new measurement matrix $M^{*}$ got by replacing the $i_2^{th}$ row of $M_1$ with the $i_2^{th}$ row of $M_2$. The $j^{th}$ column of $M^{*}$ sums to a value that is less than $\theta$. Thus, we have $f(M^{*}) \neq f(M_1)$.
\end{itemize}
Thus, the set $E$ is a valid fooling set. We now need to evaluate the size of $E$. The number of columns which sum to $\theta -1$ and $\theta$ respectively, can be evaluated by looking at the coefficients at a carefully constructed generating polynomial given by
\begin{displaymath}
(1 + Y + \ldots + Y^{m_1})(1 + Y + \ldots + Y^{m_2}) \ldots (1 + Y + \ldots + Y^{m_n}).
\end{displaymath}
This polynomial models all possible measurement vectors $(X_1, X_2, \ldots, X_n)$. Thus, we can now calculate the size of $E$ by looking at the coefficients of $Y^{\theta}$ and $Y^{\theta-1}$ in this polynomial.
\begin{eqnarray}
|E| & = & \left[Y^{\theta}\right] + \left[Y^{\theta -1}\right]\left(\prod_{i=1}^{n}(1 + Y + \ldots + Y^{m_i})\right) \\
& = & \left[Y^{\theta}\right] + \left[Y^{\theta -1}\right] \left(\prod_{i=1}^{n}\frac{1-Y^{m_i + 1}}{1 - Y}\right) \\
& = & \left[Y^{\theta}\right] + \left[Y^{\theta -1}\right] \left(\prod_{i = 1}^{n}(1 - Y^{m_i + 1}) \right)\left(\sum_{k = 1}^{\infty}\left(\begin{array}{cc} n+k-1\\  n-1\end{array}\right)Y^k \right),
\end{eqnarray}
where the last equation follows from the binomial expansion for negative exponents. Thus, we have established the required lower bound.$\Box$
\subsubsection{The $MAX$ function}
In this section, we use the tools that we have developed to study a particular example, namely the $MAX$ function. However, we no longer obtain exact results, which is to say that the single-round achievable scheme does not match the fooling set lower bound. This suggests that single round strategies are no longer optimal and it might be necessary to consider multi-round block computation strategies. Indeed, previous work in the area of communication complexity has shown a multi-round protocol that does better that our single-round scheme for the two node case. However, our proposed strategy is still exponentially better than the naive strategy of communicating all measurements. Further, it provides reasonably tight bounds and achieves the optimal scaling as the number of nodes increases.

Consider nodes $1$ through $n$ organized in a collocated network as before. For simplicity, let us suppose that for each node $i$, $X_i \in \{0, 1, \ldots, m\}$. The $MAX$ function of $n$ measurements is defined in the natural way and is denoted by $MAX_{m}(X_1, X_2, \ldots, X_n)$. We want to determine the worst case complexity of computing the $MAX$ function.
\begin{theorem}\label{thm_MAX}
The complexity of the $MAX$ function of $n$ variables from the alphabet $\{0, 1, \ldots, m\}$ is bounded as follows.
\begin{displaymath}
\log_{2}(mn + 1) \leq C(MAX_{m}(X_1, \ldots, X_n)) \leq \log_{2}\left( \begin{array}{cc} n + m  \\ m\end{array}\right).
\end{displaymath}
\end{theorem}
\textbf{Proof: } We prove the result by induction on the number of nodes $n$. For the basis step, we consider the two node problem. Consider the general achievable scheme where node $1$ sends a prefix free codeword of length $l(X_1^N)$, and node $2$ indicates its exact value for each of the instances of the block where $X_1 < X_2$. For example, if $X_1 = k$, node $2$ needs to indicate its value in the set $\{k, k+1, \ldots, m\}$. Thus, the worst case total number of bits exchanged under this scheme is given by
\begin{displaymath}
L = \max_{X_1^N}\left(l(X_1^N) +  w^{0}(X_1^N)\log_{2}(m+1) + w^{1}(X_1^N)\log_{2}m + \ldots + w^{m}(X_1^N)\log_{2}1\right ).
\end{displaymath}
Proceeding as before, we can show that, in order to ensure a valid prefix free code with codelengths $l(X_1^N)$ that satisfy Kraft inequality, the minimum $L$ is given by 
\begin{displaymath}
L = \log_{2}(m+1 + m + \ldots + 1) = N \log_{2}\left( \begin{array}{cc} m+2 \\  2 \end{array}\right).
\end{displaymath}
For the lower bound, we can verify that the set of measurement matrices with columns exclusively from the set $E$ given by
\begin{displaymath}
E = \{(0,0), (0,1), (1,0), \ldots, (0, m), (m,0)\},
\end{displaymath}
is a valid fooling set. Thus we have
\begin{displaymath}
\log_{2}(2m + 1) \leq C(MAX_{m}(X_1, X_2)) \leq \log_{2}\left( \begin{array}{cc} m+2 \\  2 \end{array}\right),
\end{displaymath}
which establishes the basis step for the induction. 

Now, let us suppose that the result is true for a network of $(n-1)$ nodes. Consider the following achievable scheme for the $n$ node network. Node $n$ transmits a prefix-free codeword of length $l(X_n^N)$. At the end of this transmission, the remaining $(n-1)$ nodes need to compute the residual $MAX$ function for each instance of the block. For example, if $X_n = k$, we are left with the task of computing the $MAX$ function of $(n-1)$ nodes on the reduced alphabet $\{k, k+1, \ldots, n\}$. Since $\{k, k+1, \ldots, n\}$ is isomorphic to $\{0, 1, \ldots , n-k\}$, this is equivalent to computing $MAX_{n-k}(X_1, \ldots , X_{n-1})$. By the induction hypothesis, there is an achievable strategy to compute this residual $MAX$ function, which we can unroll recursively. Thus the worst case total number of bits exchanged under this strategy is given by
\begin{displaymath}
L= \max_{X_n^N}\left(l(X^N_n) + w^{0}(X^N_n)C(MAX_{m}(X_1, \ldots, X_{n-1})) + \ldots + w^{m}(X^N_n)C(MAX_{0}(X_1, \ldots, X_{n-1})) \right).
\end{displaymath} 
In order to satisfy the Kraft inequality, the smallest $L$ that is feasible is given by 
\begin{eqnarray*}
L & = & N \log_{2} \sum_{i=0}^{m} 2^{C(MAX_{m-i}(X_1, \ldots, X_{n-1}))} \\
& \leq & N \log_{2} \sum_{i=0}^{m} \left( \begin{array}{cc} m+ n - i -1 \\  m-i \end{array}\right) \\
& = &  N \log_{2} \left( \begin{array}{cc} m+ n \\  m \end{array}\right)
\end{eqnarray*}
which establishes the upper bound in the induction step.

In order to prove the lower bound, we need to construct a fooling set of the appropriate size. Consider the set of $n \times N$ measurement matrices which consist of columns from the set $E$ defined by
\begin{displaymath}
E = \left\{ \left[ \begin{array}{cccc}0 \\ 0 \\ \vdots \\ 0 \end{array}\right] , \left[ \begin{array}{cccc}1 \\ 0 \\ \vdots \\ 0 \end{array}\right], \left[ \begin{array}{cccc}0 \\ 1 \\ \vdots \\ 0 \end{array}\right], \ldots , \left[ \begin{array}{cccc}0 \\ 0 \\ \vdots \\ 1 \end{array}\right], \ldots, \left[ \begin{array}{cccc}m\\ 0 \\ \vdots \\ 0 \end{array}\right], \left[ \begin{array}{cccc}0 \\ m \\ \vdots \\ 0 \end{array}\right], \ldots , \left[ \begin{array}{cccc}0 \\ 0 \\ \vdots \\ m \end{array}\right]\right\}.
\end{displaymath}
It is easy to check that this is a valid fooling set of size $(mn + 1)^N$ which gives us the required lower bound for the induction step. $\Box$

We make some observations regarding the result in Theorem \ref{thm_MAX}
\begin{itemize}
\item For fixed $m$, we have that $C(MAX_{m}(X_1, X_2, \ldots, X_n)) = \Theta(\log_{2} n)$. This agrees with the result in \cite{GiridharKumar} that the maximum rate of computing a type-threshold function is $\Theta(\frac{1}{\log_{2}n})$. Thus, the proposed achievable strategy is scaling law order-optimal. Further, we obtain better bounds on the complexity.
\begin{displaymath}
\log_{2} (mn + 1) \leq C(MAX_{m}(X_1, X_2, \ldots, X_n)) \leq \log_{2} \left( \begin{array}{cc} n + m  \\ m\end{array}\right) \leq \min( n \log_{2} (m+1), m \log_{2}(n+1)).
\end{displaymath}
\item The naive strategy for computing the $MAX$ function consists of each node communicating its measurement which has a complexity of $n \log_{2}(m+1)$. For fixed $m$, the complexity of the proposed scheme is upper bounded by $m \log_{2} (n+1)$, which is exponentially better than the naive strategy ($O(\log_{2} n)$ vs. $O(n)$).
\end{itemize}

\section{Average Case Computation of Symmetric Boolean Functions}\label{sec_coll_rand}
Consider a collocated network with nodes $1$ through $n$, where each node $i$ has a Boolean measurement $X_i \in \{0,1\}$. $X_i$ is drawn from a Bernoulli distribution with $P(X_i = 1) = : p_i$, and $\{X_i\}_{i=1}^{n}$ are independent of each other. Without loss of generality, we assume that $p_1 \leq p_2 \leq \ldots \leq p_n$. We address the following optimal distributed computation problem. \textit{Every} node wants to compute the same function $f(X_1, X_2, \ldots, X_n)$ of the measurements. Given a strategy for computing $f(X_1, X_2,$ $\ldots, X_n)$, the time of termination is a random variable. Our objective is to find communication strategies which achieve correct function computation at each node, with minimum expected total number of bits exchanged. 

In Section \ref{sec_one}, we formulate the problem of single instance computation of Boolean threshold functions. We identify a surprisingly simple policy and present a detailed proof of its optimality, by induction on the number of nodes in the network. 
In Section \ref{sec_two}, we consider the extension to the case of block computation of threshold functions, where each node has a block of measurements and we are allowed block coding. This problem is significantly harder, 
and we conjecture the structure of an optimal multi-round policy, building on the optimal policy for single instance computation. Further, we quantify the average case complexity of computing a Boolean threshold function in Section \ref{sec_avg_thres}. 

The extension of these results to an alternative model of communication, where binary information can be encoded by the presence or absence of a pulse, is studied in Section \ref{sec_three}. When considering exact computation of functions of random data, it should be noted that the time of termination is a random variable. While the optimal strategy minimizes the expected time of termination, some instances of computation might terminate earlier and some much later. In Section \ref{sec_four}, we consider the problem of approximate function computation given a fixed number of timeslots.

\subsection{Single Instance Computation of Boolean Threshold Functions}\label{sec_one}
Let us suppose each node has a single Boolean measurement and we seek to compute a given Boolean threshold function. First, we note that since each node has exactly one bit of information, it is optimal to set $b_k = X_{T_k}$. Indeed, for any other choice $b_k' = g(b_1, \ldots, b_{k-1}, X_{T_{k}})$, the remaining nodes can reconstruct $b_k'$ since they already know $b_i, \ldots, b_{k-1}$. Thus the only freedom available is in choosing the transmitting node $T_k$ as a function of $b_1, b_2, \ldots, b_{k-1}$, for otherwise the transmission itself could be avoided. We call this the \textit{ordering problem}. Thus, by definition, the order can dynamically depend on the previous broadcast bits. In this paper, we address the ordering problem for a class of Boolean functions, namely threshold functions.

We will denote the set of measurements of nodes $1$ through $n$ by $(X_1, X_2, \ldots, X_n)$ which is abbreviated as $\mathbf{X}^{n}$. We will use $\mathbf{X}^{n}_{-i}$ to denote the set of measurements $(X_1, \ldots, X_{i-1},$ $X_{i+1}, \ldots, X_n)$. As a natural extension, we use $\mathbf{X}^{n}_{-( i,j)}$ to denote the set of measurements $(X_1, \ldots,$ $X_{i-1}, X_{i+1}, \ldots, X_{j-1}, X_{j+1}, \ldots, X_n)$, where $i < j$.

\begin{definition}[Boolean threshold functions]
A Boolean threshold function $\Pi_{\theta}(X_1, X_2, \ldots, X_n)$ is defined as
\begin{displaymath}
\Pi_{\theta}(X_1, X_2, \ldots, X_n) = \left\{ \begin{array}{l} 1 \quad \textrm{if } \sum_{i}X_i \geq \theta , \\ 0 \quad \textrm{otherwise.}\end{array}\right.
\end{displaymath}
\end{definition}

The class of threshold functions has the property that, if one of the nodes' measurements is known, the residual function is still a threshold function. Given a function $\Pi_{n-k}(\mathbf{X}^{n})$, if node $i$ transmits its bit, we are left with the residual task of computing $\Pi_{n-k-1}(\mathbf{X}^n_{-i})$ if $X_i =1$, and $\Pi_{n-k}(\mathbf{X}^n_{-i})$ if $X_i = 0$.  Thus, the ordering problem can be formulated as a dynamic programming problem. Let $C(\Pi_{n-k}(\mathbf{X}^{n}))$ denote the minimum expected number of bits required to compute $\Pi_{n-k}(\mathbf{X}^{n})$. The dynamic programming equation is
\begin{equation} \label{dyn_prog_eqn}
C(\Pi_{n-k}(\mathbf{X}^{n})) = \min_{i}\{1 + p_i C(\Pi_{n-k-1}(\mathbf{X}^{n}_{-i})) + (1-p_i)C(\Pi_{n-k}(\mathbf{X}^{n}_{-i}))\}.
\end{equation}
with boundary condition $C(\Pi_{a}(\mathbf{X}^m)) = 0$ if $a = 0$ or $a > m$.

To begin with, we argue that solving (\ref{dyn_prog_eqn}) for each $n$ and $k$ does indeed yield the optimal strategy for computing Boolean threshold functions. In particular, to derive the optimal strategy for computing $\Pi_{n-k}(\mathbf{X}^n)$, we first determine which node must transmit first, by solving (\ref{dyn_prog_eqn}) for $n, k$. Then, depending on whether $X_{T(1)} = 0$ or $X_{T(1)} = 1$, we are left with the residual task of computing $\Pi_{n-k}(\mathbf{X}^{n}_{-T(1)})$ or $\Pi_{n-k-1}(\mathbf{X}^{n}_{-T(1)})$. We can determine which node should transmit next in either case, from the solution of (\ref{dyn_prog_eqn}) for $n-1, k-1$ or $n-1, k$ respectively. Proceeding recursively, one can unroll the optimal strategy for computing $\Pi_{n-k}(X_1, X_2, \ldots X_n)$.

In (\ref{dyn_prog_eqn}), we recognise that the single-stage cost is uniformly $1$. More generally, given a function $f(\cdot): [0,1] \rightarrow \mathbf{R}^{+}$, one can write down a more general dynamic programming equation. 
\begin{equation} \label{dyn_prog_eqn_gen}
C(\Pi_{n-k}(\mathbf{X}^{n})) = \min_{i}\{f(p_i) + p_i C(\Pi_{n-k-1}(\mathbf{X}^{n}_{-i})) + (1-p_i)C(\Pi_{n-k}(\mathbf{X}^{n}_{-i}))\}.
\end{equation}
Here, one can view $f(p_i)$ as the cost of communicating the information of node $i$ which has $P(X_i = 1) = p_i$. Indeed, for the case of single instance computation, we have $f(p) \equiv 1$. In the sequel, we will see how this general dynamic programming formulation will allow us to study other problems of interest.

For general $f(\cdot)$, solving the dynamic programing equation (\ref{dyn_prog_eqn_gen}) may be intractable. Further, it is unclear at the outset if the optimal strategy will depend only on the ordering of the $p_i$s, or their particular values. This makes the explicit solution of (\ref{dyn_prog_eqn_gen}), or even (\ref{dyn_prog_eqn}), for all $n$, $k$ and $(p_1, p_2, \ldots p_n)$ notoriously hard. However, under some conditions on $f(\cdot)$, we can derive a very simple characterization of the optimal strategy for each $n$ and $0 \leq k \leq n-1$. Further, we observe that optimal strategy is independent of the particular values of the $p_i$s, but only depends on their relative ordering. 
\begin{lemma}\label{thm_seq_gen}
Let $f(\cdot): [0,1] \rightarrow \mathbf{R}^{+}$ be a function such that 
\begin{itemize}
\item $f(p) = f(1-p)$.
\item $\frac{f(p)}{p}$ is a monotone non-increasing function of $p$. 
\end{itemize}
Then the minimum in (\ref{dyn_prog_eqn_gen}) is attained by $k+1$. That is,
\begin{equation}\label{eqn_argmin}
k+1 \in \underset{i}{\operatorname{argmin}}\left\{f(p_i) + p_i C(\Pi_{n-k-1}(\mathbf{X}^{n}_{-i})) + (1-p_i)C(\Pi_{n-k}(\mathbf{X}^{n}_{-i})\right\}.
\end{equation}
This result is true for all $n$ and all $0 \leq k \leq n-1$ and all probability distributions with $p_1 \leq p_2 \leq \ldots \leq p_n$.
\end{lemma}
\textbf{Proof:} We define the following expressions.
\begin{multline}
T_{m,k,i}(\mathbf{X}^{m}) = p_{k+1}C(\Pi_{m-k-1}(\mathbf{X}^m_{-(k+1)})  + (1-p_{k+1})C(\Pi_{m-k}(\mathbf{X}^m_{-(k+1)}) \\
-  p_{i}C(\Pi_{m-k-1}(\mathbf{X}^m_{-i}) - (1-p_i) C(\Pi_{m-k}(\mathbf{X}^m_{-i}) \nonumber
\end{multline}
\begin{multline*}
S^{(1)}_{m,k,i}(\mathbf{X}^{m}) := (p_{k+1} - p_{i})C(\Pi_{m-k-1}(\mathbf{X}^{m}_{-(k+1, i)})) + (1-p_{k+1})C(\Pi_{m-k}(\mathbf{X}^{m}_{-(k+1)})) \\
- (1-p_{i})C(\Pi_{m-k}(\mathbf{X}^{m}_{-i})). 
\end{multline*}
\begin{equation*} 
S^{(2)}_{m,k,i}(\mathbf{X}^{m}) := (p_{i} - p_{k+1})C(\Pi_{m-k-1}(\mathbf{X}^{m}_{-(i, k+1)})) + p_{k+1}C(\Pi_{m-k-1}(\mathbf{X}^{m}_{-(k+1)})) - p_{i}C(\Pi_{m-k-1}(\mathbf{X}^{m}_{-i})).
\end{equation*}

We establish the above theorem by induction on the number of nodes $n$. However, we need to load the induction hypothesis. Consider the following induction hypothesis.
\begin{eqnarray}
\textrm{(a) } T_{m,k,i}(\mathbf{X}^{m}) & \leq & f(p_i) - f(p_{k+1}) \quad \textrm{for all } 0 \leq k \leq (m-1), 1 \leq i \leq m  \nonumber\\
\textrm{(b) } S^{(1)}_{m,k,i}(\mathbf{X}^{m}) & \leq & (1-p_{k+1})f(p_i) - (1- p_i)f(p_{k+1}) \quad \textrm{for all } 0 \leq k+1 \leq (m-1), k+2 \leq i \leq m  \nonumber \\
\textrm{(c) } S^{(2)}_{m,k,i}(\mathbf{X}^{m}) & \leq & p_{k+1}f(p_i) - p_if(p_{k+1}) \quad \textrm{for all } 0 \leq k \leq (m-1), 1 \leq i < k+1  \nonumber
\end{eqnarray}
Observe that part $(a)$ immediately establishes (\ref{eqn_argmin}).

The basis step for $m = 1$ is trivially true. Let us suppose the induction hypothesis is true for all $m \leq n$. We now proceed to prove the hypothesis for $m = n+1$. 
\begin{lemma}\label{lemma_three}
For fixed $k$ and $i \geq k+2$, we have 
\begin{displaymath}
S^{(1)}_{n+1,k,i}(\mathbf{X}^{n+1}) \leq (1-p_{k+1})f(p_i) - (1- p_i)f(p_{k+1}).
\end{displaymath}
\end{lemma}
\textbf{Proof:} See Appendix \ref{sec_proof_three}.
\begin{lemma}\label{lemma_four} 
For fixed $k$ and $i \leq k$, we have 
\begin{displaymath}
S^{(2)}_{n+1,k,i}(\mathbf{X}^{n+1}) \leq p_{k+1}f(p_i) - p_if(p_{k+1}).
\end{displaymath}
\end{lemma}
\textbf{Proof:} See Appendix \ref{sec_proof_four}.

Lemmas \ref{lemma_three} and \ref{lemma_four} establish the induction step for parts $(b)$ and $(c)$ of the induction hypothesis. We now proceed to show the induction step for part $(a)$.
\begin{lemma}\label{lemma_one}
For fixed $k$ and $i \geq k+2$, we have 
\begin{displaymath}
T_{n+1, k, i}(\mathbf{X}^{n+1}) \leq S^{(1)}_{n+1,k,i}(\mathbf{X}^{n+1}) + p_{k+1} f(p_i) - p_{i} f(p_{k+1}).
\end{displaymath}
\end{lemma}
\textbf{Proof:} See Appendix \ref{sec_proof_one}.
\begin{lemma}\label{lemma_two}
For fixed $k$ and $i \leq k$, we have 
\begin{displaymath}
T_{n+1, k, i}(\mathbf{X}^{n+1}) \leq S^{(2)}_{n+1, k, i}(\mathbf{X}^{n+1}) + (1-p_{k+1})f(p_i) - (1-p_i) f(p_{k+1}).
\end{displaymath}
\end{lemma}
\textbf{Proof:} See Appendix \ref{sec_proof_two}.

Applying Lemmas \ref{lemma_one} and \ref{lemma_two} together with Lemmas \ref{lemma_three} and \ref{lemma_four}, we see that $T_{n+1, k, i}(\mathbf{X}^{n+1}) \leq 0$ for all $0 \leq k \leq n$ and $i \neq k+1$. For the case $i = k+1$, we have $T(n+1, k, k+1) = 0$ trivially. This completes the induction step for part $(a)$, and the proof of the Theorem. $\Box$

Using Lemma \ref{thm_seq_gen}, we can now simply derive the optimal sequential communication strategy for computing a single instance of the Boolean threshold function $\Pi_{n-k}(\mathbf{X}^n)$. 
\begin{theorem}\label{thm_seq_bool_threshold}
In order to compute a single instance of the Boolean threshold function $\Pi_{n-k}(\mathbf{X}^n)$, it is optimal for node $(k+1)$ to transmit its bit first. 
\end{theorem}
\textbf{Proof: }In the case of single instance computation, we have $f(p) \equiv 1$. Hence, trivially, we have that $f(p) = f(1-p)$, and that $\frac{f(p)}{p}$ is a monotone non-increasing function of $p$. From Lemma \ref{thm_seq_gen}, we have 
\begin{displaymath}
k+1 \in \underset{i}{\operatorname{argmin}}\left\{f(p_i) + p_i C(\Pi_{n-k-1}(\mathbf{X}^{n}_{-i})) + (1-p_i)C(\Pi_{n-k}(\mathbf{X}^{n}_{-i})\right\}.
\end{displaymath}
Thus, in order to compute the Boolean threshold function $\Pi_{n-k}(\mathbf{X}^n)$, it is optimal for node $k+1$ to transmit first. $\Box$
\begin{remark}
At the outset, there are two heuristics that one may apply to the ordering problem. First, if we believe that $\Pi_{n-k}(\mathbf{X}^n)$ evaluates to $0$, the conditional optimal strategy is for nodes to transmit in order starting with node $1$. Alternately, if we believe that $\Pi_{n-k}(\mathbf{X}^n)$ evaluates to $1$, the conditional optimal strategy is for nodes to transmit in reverse order starting with node $n$. Thus, the result in Theorem \ref{thm_seq_bool_threshold} can be viewed as an appropriate \textit{hedging} solution which safeguards against the event that $\Pi_{n-k}(\mathbf{X}^n)$ could evaluate to $0$ or $1$. It is indeed surprising that a particularly simple hedging strategy is optimal for all $n$, all $k$ and all probability distributions, and that it does not depend on the actual values of the probabilities but only on their order.
\end{remark}
\subsection{Block Computation of Boolean Threshold Functions}\label{sec_two}
We now shift attention to the case where we allow nodes to accumulate a block of $N$ measurements, and thus achieve improved efficiency by using block codes. The most general class of interactive strategies are those where the identity of the node transmitting the $k^{th}$ bit, say $T_k$ can depend arbitrarily on all previously broadcast bits, and the $k^{th}$ bit itself can depend arbitrarily on all previously broadcast bits as well as $T_k$'s block of measurements. We require that all nodes compute the function with zero error for the block, and wish to minimize the expected number of bits exchanged per instance of computation, denoted $\mathcal{C}(\Pi_{n-k}(\mathbf{X}^n))$. While the problem of finding the optimal strategy in this general class of strategies appears intractable, we derive the optimal solution under a restricted class of strategies. The restriction we impose is natural, and we conjecture that the optimal strategy in this restricted class is also optimal among all interactive strategies.

Define the following restricted class of \emph{coherent strategies}. 
\begin{definition}{Coherent Strategies}
When computing $\Pi_{n-k}(\mathbf{X}^{n})$ for a block of $N$ measurements, a coherent strategy mandates that the first node to transmit, say node $T(1)$, must declare its entire block using a Huffman code. Note that this does not exclude interactive strategies, since, subsequent to node $T(1)$'s  transmission, we have two subproblems over sub-blocks of measurements corresponding to instances where $X_{T(1)} = 0$ and $X_{T(1)} =1$. For each of these subproblems, we could potentially have different nodes transmitting first. Thus nodes may transmit more than once. However each of these nodes are again constrained to communicate their entire subblock of measurements. 
\end{definition}
\begin{theorem}\label{thm_seq_bool_threshold_block}
In the restricted class of coherent strategies, in order to compute $\Pi_{n-k}(\mathbf{X}^{n})$ for a block of measurements, it is optimal for node $k+1$ to transmit its entire block first, using the Huffman code. This result is true for asymptotically long block lengths, for all $n$, all $0 \leq k \leq n-1$, and all probability distributions with $p_1 \leq p_2 \leq \ldots \leq p_n$.
\end{theorem}
\textbf{Proof: } Let us suppose node $i$ transmits first. Under a coherent strategy, node $i$ must communicate its entire block, which requires an average description length of $H(p_i)$ bits per instance. This can be achieved asymptotically by using the Huffman code to compress node $i$'s block of measurements \footnote{For clarity of presentation, we will ignore the fact that the Huffman code for block length $N$ has average codelength between $\lfloor NH(p)\rfloor$ and $\lfloor NH(p)\rfloor + 1$ bits. The extra one bit can be amortized over long block lengths.}. Subsequent to node $i$'s transmission, we are left with the residual tasks of computing $\Pi_{n-k-1}(\mathbf{X}^{n}_{-i})$ for the subblock where $X_i = 1$, and $\Pi_{n-k}(\mathbf{X}^{n}_{-i})$ for the subblock where $X_i =0$. These are two block computation problems again. Let $\mathcal{C}_{U}(\Pi_{n-k}(\mathbf{X}^n)$ denote the minimum number of bits per instance, that must be exchanged under this restricted class of strategies. We can write a dynamic programming equation as before.
\begin{equation} \label{dyn_prog_eqn_block}
\mathcal{C}_{U}(\Pi_{n-k}(\mathbf{X}^{n})) = \min_{i}\{H(p_i) + p_i \mathcal{C}_{U}(\Pi_{n-k-1}(\mathbf{X}^{n}_{-i})) + (1-p_i)\mathcal{C}_{U}(\Pi_{n-k}(\mathbf{X}^{n}_{-i}))\},
\end{equation}
where $H(p)$ is the standard binary entropy function defined by $H(p) = -p\log_{2}(p) - (1-p)\log_{2}(1-p)$. The boundary condition for (\ref{dyn_prog_eqn_gen}) is given by $\mathcal{C}_{U}(\Pi_{a}(\mathbf{X}^m)) = 0$ if $a=0$ or $a > m$. 

Observe that (\ref{dyn_prog_eqn_block}) is a special case of (\ref{dyn_prog_eqn_gen}) where $f(p) = H(p)$. Thus, for the class of coherent strategies, the problem of optimal computation once again reduces to an ordering problem. If we can show that $H(p)$ satisfies the conditions in Lemma \ref{thm_seq_gen}, the result follows immediately. Clearly $H(p) = H(1-p)$ and one can verify that 
\begin{displaymath}
\frac{d \left(\frac{H(p)}{p}\right)}{dp} = \frac{\log_2 (1-p)}{p^2} \leq 0.
\end{displaymath}
Thus, we have that $\frac{H(p)}{p}$ is a non-decreasing function of $p$. Hence, from Lemma \ref{thm_seq_gen}, the optimal strategy for computing $\Pi_{n-k}(\mathbf{X}^{n})$ for a block of measurements is for node $k+1$ to transmit its entire block first, using the Huffman code. $\Box$

\begin{remark}
The proposed optimal strategy is inherently interactive, since nodes do transmit more than once. This is due to the recursive splitting of the original block of measurements depending on nodes' transmissions. This is illustrated in the computation tree for $\Pi_{2}(\mathbf{X}^3)$, where node $2$ first transmits its entire block using a Huffman code, and the computation proceeds as shown. In practice, all nodes need to agree \textit{a priori} on a traversal order in the computation tree, e.g., depth-first traversal or breadth-first traversal.
\end{remark}

\begin{remark}
The proposed optimal strategy is asymptotically optimal in the limit of long blocks. This is necessary to achieve an average description length of $H(p_i)$ bits per instance. In practice, one could simply choose a large enough block length $N$ so that the average description length is close enough to the entropy. In this context, it is important to note that, as the computation proceeds, the original block gets recursively subdivided into smaller and smaller subblocks of measurements. Each of these subblocks needs to be large enough to achieve an average description length that is close enough to the entropy of the transmitting node. Thus, in the worst case, we could have upto $2^n$ subblocks in the computation tree, and we assume that each of these subblocks are large enough, which is ensured by choosing $N$ to be suitably large. 
\end{remark}

\subsubsection{A Strategy-independent Lower Bound}
Next, we would like to determine if the class of coherent strategies considered above is rich enough to include the absolute optimal strategy for interactive block computation without any restrictions on a node encoding all its information using a Huffman code. Intuitively, since all the instances of the block are independent and identically distributed, it appears suboptimal for nodes to communicate only partial information regarding their blocks at any stage. Thus, it is plausible that, under the optimal strategy, one node communicates its entire block, and the computation proceeds recursively from there. However, establishing this fact rigorously is a formidable challenge. In this subsection, we describe a possible approach toward establishing this result, by adapting the concept of \textit{fooling sets}. Fooling sets are a classical tool for establishing lower bounds in communication complexity \cite{KushiNisan}, and have recently been used to establish tight lower bounds on the minimum number of bits exchanged in the worst-case in collocated networks \cite{KowshikKumar_ITW}, and tree networks \cite{KowshikKumar_Tree}. We describe an extension of fooling sets to the probabilistic scenario and use this to establish a lower bound. 


We recall that, for the threshold function $\Pi_{n-k}(\mathbf{X}^n)$, a valid fooling set of maximum size is given by 
\begin{displaymath}
E_{n,n-k} := \{\mathbf{X}^n: \sum_{i=1}^{n}X_i = n - k \textrm{ or } \sum_{i=1}^{n}X_i = n - k - 1\}
\end{displaymath}
Any correct protocol for distributed computation of $\Pi_{n-k}(\mathbf{X}^n)$ partitions the function matrix into monochromatic rectangles \cite{KushiNisan}. Further, each rectangle in the partition can contain at most one element of $E_{n,n-k}$. Let $D(\Pi_{n-k}(\mathbf{X}^n))$ be the set of all protocol partitions of the function matrix of $\Pi_{n-k}(\mathbf{X}^n)$ which respect the fooling set constraints. Suppose we use a protocol with associated partition $d$, the number of bits that must be exchanged under this protocol is lower bounded by the entropy of this partition, denoted by $H(p(d))$, where $p(d)$ is the implied probability distribution on the elements of the partition. Thus, the number of bits that must be exchanged under \textit{any} protocol is bounded by
\begin{equation}\label{block_lower_bound}
\mathcal{C}(\Pi_{n-k}(\mathbf{X}^n)) \geq \min_{d \in D(\Pi_{n-k}(\mathbf{X}^n))}H(p(d)) =: \mathcal{C}_{L}(\Pi_{n-k}(\mathbf{X}^n)).
\end{equation}
We conjecture that this lower bound is achievable and in particular, the optimal strategy described in Theorem \ref{thm_seq_bool_threshold_block} achieves it.
\begin{conjecture}
The lower bound described in (\ref{block_lower_bound}) satisfies the dynamic programming equation in (\ref{dyn_prog_eqn_block}). 
\begin{displaymath}
\mathcal{C}_{L}(\Pi_{n-k}(\mathbf{X}^{n})) = \min_{i}\{H(p_i) + p_i \mathcal{C}_{L}(\Pi_{n-k-1}(\mathbf{X}^{n}_{-i})) + (1-p_i)\mathcal{C}_{L}(\Pi_{n-k}(\mathbf{X}^{n}_{-i}))\}.
\end{displaymath}
Since $\mathcal{C}_{L}(\Pi_{n-k}(\mathbf{X}^{n})) \leq \mathcal{C}(\Pi_{n-k}(\mathbf{X}^{n})) \leq \mathcal{C}_{U}(\Pi_{n-k}(\mathbf{X}^{n}))$, we conjecture that the optimal strategy described in Theorem \ref{thm_seq_bool_threshold_block} is in fact optimal among all interactive strategies.
\end{conjecture}

We note that the above conjecture has been verified by hand for all threshold functions of three variables. A formal proof of the conjecture, however, remains a challenge for the future. 

\subsubsection{Average Case Complexity of Computing Boolean Threshold Functions} \label{sec_avg_thres}
In this section, we quantify the average case complexity of computing Boolean threshold functions in collocated networks. For simplicity, we suppose that nodes' measurements are independent and identically distributed, and propose a natural block computation strategy that is easy to analyze.
\begin{theorem}\label{thm_bool_avg}
Suppose that the nodes' measurements $X_{1}, X_{2}, \ldots, X_{n}$ are independent and identically distributed with $p(X_{i} = 1) = p$. Then, the average case complexity of zero error block computation of the threshold function $\Pi_{\theta}(X_1, X_2, \ldots, X_n)$ is $O(\theta)$ bits.
\end{theorem}
\textbf{Proof:} We need to describe a coding strategy which achieves zero error block computation, as block length $N$ goes to infinity. Let us suppose that nodes communicate in reverse order starting with node $n$. Node $n$ encodes its block of $N$ measurements using a Huffman code which requires $\lceil NH(p)\rceil$ bits. Having heard all previous transmissions, each successive node discards the instances of the block that are already \textit{determined}, i.e., those instances of the block that have already recorded $\theta$ ones. It then constructs the Huffman code for the remaining instances of the block. Let the number of determined instances after node $i+1$ transmits be denoted by random variable $Z_i$. Then, the average complexity of computing a function block of length $N$ is given by
\begin{equation}
\sum_{i = 1}^{n} (N - \mathbf{E}(Z_i))H(p) = \theta NH(p) + NH(p)\sum_{i = \theta}^{n-1}\sum_{j = 0}^{\theta -1}\begin{pmatrix}i \\ j\end{pmatrix} p^{j}(1-p)^{i-j}. \label{coll_eqn_1} 
\end{equation}
We need to somehow carefully upper bound the RHS in the (\ref{coll_eqn_1}). We start by establishing the following lemma.
\begin{lemma}\label{taylor_lemma}
Define $g_{\theta} := \frac{x^{\theta}}{1-x}$ for $\theta$ a positive integer. Then
\begin{displaymath}
 f_{\theta}^{(\theta -1)} := \frac{d^{(\theta -1)}g_{\theta}}{d x^{(\theta -1)}} = (\theta - 1)!\left(\frac{1}{(1-x)^{\theta}} - 1\right).
\end{displaymath}
\end{lemma}
\textbf{Proof of Lemma: } The proof is by induction on $\theta$. For $\theta = 1$, we have $g_1 = \frac{x}{1-x} = g_1^{(0)}$ trivially. For $\theta>1$, observe that $g_{\theta -1} - g_{\theta} = x^{\theta -1}$ and hence $g_{\theta}^{(\theta -1)} = g_{\theta -1}^{(\theta-1)}-(\theta -1)!$. By the induction assumption, we have
\begin{displaymath}
g_{\theta}^{(\theta -1)} = \frac{d}{dx}\left(\frac{(\theta -2)!}{(1-x)^{\theta -1}}\right) - (\theta -1)! = \left(\frac{(\theta -1)!}{(1-x)^{\theta}} - 1\right),
\end{displaymath}
which completes the induction. $\Box$

We now proceed to show that the second term on the RHS in (\ref{coll_eqn_1}) is smaller than $\theta NH(p)\left(\frac{1-p}{p}\right)$ for each $n$. The proof is by induction on $\theta$. For $\theta=1$, the second term is given by 
\begin{displaymath}
\sum_{i=1}^{n-1}NH(p)p^{\theta}(1-p)^{i} = NH(p)\frac{(1-p) - (1-p)^{n}}{p} < \frac{NH(p)(1-p)}{p}. 
\end{displaymath}
Define $R_{\theta}^{n} := \sum_{i=\theta}^{n-1}\sum_{j = 0}^{\theta -1} \begin{pmatrix}i \\ j\end{pmatrix} p^{j}(1-p)^{i-j}$. Then, we have the following recursion: 
\begin{displaymath}
R_{\theta}^{n} = R_{\theta - 1}^{n} + \sum_{i = \theta}^{n-1}\begin{pmatrix}i \\ \theta -1\end{pmatrix}p^{\theta -1}(1-p)^{i-\theta + 1} - \sum_{j=0}^{\theta -2}\begin{pmatrix}\theta -1\\ j\end{pmatrix}p^{j}(1-p)^{\theta -1 -j}.
\end{displaymath}

From the induction hypothesis, we have that
\begin{eqnarray}
R_{\theta}^{n} & \leq & \frac{(\theta -1)(1-p)}{p} + \left(\sum_{i = \theta}^{n-1}\begin{pmatrix}i \\ \theta-1 \end{pmatrix}p^{\theta -1}(1-p)^{i-\theta + 1}\right) - 1 + p^{\theta -1} \nonumber \\
& \leq &  \frac{(\theta -1)(1-p)}{p} + \left(\frac{p^{\theta -1}}{(\theta-1)!}\sum_{i = \theta}^{\infty}i(i-1)\ldots(i-\theta +2)(1-p)^{i-\theta + 1}\right) -1 + p^{\theta -1} \nonumber \\
& =  & \frac{(\theta -1)(1-p)}{p} + \frac{p^{\theta -1}}{(\theta-1)!}\frac{d^{(\theta -1)}}{d x^{(\theta -1)}}\left(\frac{x^{\theta}}{1-x}\right) - 1 + p^{\theta -1}. \label{intermed_1}
\end{eqnarray}
Now, applying Lemma \ref{taylor_lemma} in (\ref{intermed_1}), we can show $R_{\theta}^{n} \leq   \frac{\theta(1-p)}{p}$, which establishes the induction step. Substituting the upper bound for the second term in the RHS of (\ref{coll_eqn_1}), we obtain that the total number of bits transmitted is less than $\frac{\theta NH(p)}{p}$ for all $n$. This yields a sum rate of $\frac{\theta H(p)}{p}$ which completes the proof. $\Box$

We make some observations regarding the above result. 
\begin{itemize}
\item[(i)] For a type-threshold function \cite{GiridharKumar} with threshold vector $[\theta_1, \theta_2]$, we can run two parallel schemes with thresholds $[\theta_1, 0]$ and $[0, \theta_2]$, thus attaining a sum rate $\frac{(\theta_1 + \theta_2)H(p)}{p}$. Since we typically consider $\theta_1, \theta_2$ to be constants independent of $n$, we obtain that the average case complexity of computing Boolean threshold functions is $O(1)$.
\item[(ii)] As a special case, the average case complexity of computing a symmetric Boolean Disjunctive Normal Form with bounded minterms is $\Theta(1)$.  
\end{itemize}
\subsection{Computation under an alternate communication model}\label{sec_three}
In this section, we illustrate how we can adapt the solution to the general dynamic programming equation described in Lemma \ref{thm_seq_gen} to a different communication model. We return to the problem of computing a single instance of a Boolean threshold function $\Pi_{n-k}(\mathbf{X}^n)$ in the broadcast scenario. Let us suppose that time is slotted, and that nodes transmit information in the form of pulses, which have unit energy cost. Under this alternate communication model, our modified objective is to minimize the expected total energy expended in transmissions. 

In contrast to Section \ref{sec_one} where the cost of transmitting a bit is uniformly $1$, under the pulse model of communication, silence can be used to convey information with zero cost. This is similar to the silence-based communication model studied in \cite{DhulipalaFragouliOrlitsky}. Thus, the communication problem is no longer trivial. However, since each node makes a Boolean measurement, the value of its bit can be mapped to the presence or absence of a pulse in two ways. Either node $i$ transmits a pulse to indicate $X_i = 1$ and remains silent to indicate $X_i = 0$, or vice versa. Clearly, the optimal communication strategy is as follows:
\begin{itemize}
\item If $p_i \leq \frac{1}{2}$, then node $i$ transmits a pulse to indicate $X_i = 1$.
\item If $p_i \geq \frac{1}{2}$, then node $i$ transmits a pulse to indicate $X_i = 0$.
\end{itemize}
We are still left with the problem of determining the optimal ordering of transmissions.

Let $\mathcal{E}(\Pi_{n-k}(\mathbf{X}^n))$ be the minimum expected total energy required in order to compute the threshold function $\Pi_{n-k}(\mathbf{X}^n)$ under this communication model. The problem of minimizing the expected total energy can be formulated as a dynamic programming equation as follows
\begin{equation}\label{dyn_prog_alt_comm}
\mathcal{E}(\Pi_{n-k}(\mathbf{X}^{n})) = \min_{i}\{\min(p_i, 1-p_i) + p_i \mathcal{E}(\Pi_{n-k-1}(\mathbf{X}^{n}_{-i})) + (1-p_i)\mathcal{E}(\Pi_{n-k}(\mathbf{X}^{n}_{-i}))\}
\end{equation}
From Lemma \ref{thm_seq_gen}, we have the following result.
\begin{theorem}
In order to compute a single instance of the Boolean threshold function $\Pi_{n-k}(\mathbf{X}^n)$ under the pulse communication model, it is optimal for node $k+1$ to transmit first.
\end{theorem}
\textbf{Proof: }Observe that (\ref{dyn_prog_alt_comm}) is a special case of (\ref{dyn_prog_eqn_gen}) where $f(p) = \min(p, 1-p)$. Hence, in order to establish the result, it is sufficient to show that $\min(p, 1-p)$ satisfies the conditions in Lemma \ref{thm_seq_gen}. Indeed, $\min(p,1-p)$ is symmetric about $p = \frac{1}{2}$ and we have,
\begin{displaymath}
g(p) = \frac{\min(p, 1-p)}{p} = \left \{ \begin{array}{ll} 1 \quad \textrm{if } p \leq \frac{1}{2}, \\ \frac{1-p}{p} \quad \textrm{if } p > \frac{1}{2.}\end{array}\right.
\end{displaymath} 
Thus, $\frac{\min(p, 1-p)}{p}$ is a monotone non-increasing function of $p$. The theorem then follows directly from Lemma \ref{thm_seq_gen}. $\Box$
\subsection{Approximate Function Computation}\label{sec_four}
In Sections \ref{sec_one} through \ref{sec_three}, we have considered the problem of computing Boolean threshold functions with zero error. While we have focused on constructing optimal strategies to minimize the expected total number of bits exchanged during computation, we must note that the worst-case total number of bits exchanged might still be $n$. In some applications however, we might have a constraint on the number of bits exchanged, or equivalently, the number of timeslots available for computation. In this case, one cannot always hope to compute the function exactly. Instead, we consider \textit{approximate} function computation where we seek to minimize certain error metrics. 

To begin with, let us consider the class of Boolean threshold functions. As before, we permit all interactive strategies where the choice of next transmitting node can depend arbitrarily on all previously broadcast bits. Let us suppose that we are allowed to exchange at most $(n-\theta)$ bits in order to compute the threshold function $\Pi_{n-k}(\mathbf{X}^n)$. We propose two metrics of error, namely probability of error and conditional entropy of the function.
\begin{itemize}
\item \textbf{Probability of error: } Let $P_e^{(\theta)}(\Pi_{n-k}(\mathbf{X}^n))$ denote the minimum probability of error after at most $(n-\theta)$ bits are exchanged. Note that these bits are exchanged in sequential fashion, since we are computing in a broadcast network. Hence, the identity of each successive transmitting node can depend on the previously transmitted bits. The sequential nature of this problem permits a dynamic programming formulation analogous to (\ref{dyn_prog_eqn_gen}). 
\begin{equation}\label{dyn_prog_proberror}
P_e^{(\theta)}(\Pi_{n-k}(\mathbf{X}^n)) = \min_{i} \{p_i P_e^{(\theta)}(\Pi_{n-k-1}(\mathbf{X}^n_{-i}) + (1-p_i) P_e^{(\theta)}(\Pi_{n-k}(\mathbf{X}^n_{-i}) \},
\end{equation}
with the boundary condition $P_e^{(\theta)}(\Pi_{\theta-k}(\mathbf{X}^{\theta})) = \min (P(\Pi_{\theta-k}(\mathbf{X}^{\theta}) = 1), P(\Pi_{\theta-k}(\mathbf{X}^{\theta}) = 0))$.  
\item \textbf{Conditional entropy of function: } Let $H^{(\theta)}(\Pi_{n-k}(\mathbf{X}^n))$ denote the minimum conditional entropy of the function after at most $(n-\theta)$ bits are exchanged. As before, the identity of each successive transmitting node can depend on the previously transmitted bits. Once again, the sequential nature of this problem permits a dynamic programming formulation analogous to (\ref{dyn_prog_eqn_gen}). 
\begin{equation}\label{dyn_prog_entropy}
H^{(\theta)}(\Pi_{n-k}(\mathbf{X}^n)) = \min_{i} \{p_i H^{(\theta)}(\Pi_{n-k-1}(\mathbf{X}^n_{-i}) + (1-p_i) H^{(\theta)}(\Pi_{n-k}(\mathbf{X}^n_{-i}) \},
\end{equation}
with the boundary condition $H^{(\theta)}(\Pi_{\theta-k}(\mathbf{X}^{\theta})) = H(\Pi_{\theta-k}(\mathbf{X}^{\theta}))$. 
\end{itemize}

\subsubsection{Counter-example}\label{counter_eg}
At fsubirst glance, one would expect that the optimal strategy for approximate function computation would match the strategy for exact function computation, thus verifying that the strategy proposed in Theorem \ref{thm_seq_bool_threshold} is \textit{increasingly correct}. Unfortunately, this is not true as shown by the following counter example. 

Let us suppose that we want to compute $\Pi_{2}(X_1, X_2, X_3)$ and we are allowed to exchange exactly one bit. We have exactly three choices of strategy. Either node $1$ transmits first, or node $2$, or node $3$. Consider the case where $p_1 = 0.7, p_2 = 0.82, p_3 = 0.84$, then one can calculate the conditional entropy when node $1$ transmits first (respectively node $2$ and node $3$).
\begin{eqnarray*}
H^{(2)}(\Pi_{2}(X_1, X_2, X_3)| X_1) & = & p_1 H((1-p_2)(1-p_3)) + (1-p_1)H(p_2 p_3) = 0.4002.\\
H^{(2)}(\Pi_{2}(X_1, X_2, X_3)| X_2) & = & p_2 H((1-p_1)(1-p_3)) + (1-p_2)H(p_1 p_3) = 0.4991.\\
H^{(2)}(\Pi_{2}(X_1, X_2, X_3)| X_3) & = & p_3 H((1-p_1)(1-p_2)) + (1-p_3)H(p_1 p_2) = 0.4121.
\end{eqnarray*}
Contrary to our expectation, it is not always optimal for node $2$ to transmit first. This is also true for the probability of error metric. Again, consider the approximate computation of $\Pi_{2}(X_1, X_2, X_3)$ where we are only allowed to exchange exactly one bit. For the case where $p_1 = 0.6, p_2 = 0.72, p_3 = 0.84$, the probability of error expressions for the three strategies are given by
\begin{eqnarray*}
P_e^{(2)}(\Pi_{2}(X_1, X_2, X_3)| X_1) & = & p_1 \min ((1-p_2)(1-p_3), 1 - (1-p_2)(1-p_3)) \\
& & + (1-p_1)\min (p_2 p_3, 1 - p_2 p_3) = 0.1850, \\
P_e^{(2)}(\Pi_{2}(X_1, X_2, X_3)| X_2) & = & p_2 \min ((1-p_1)(1-p_3), 1 - (1-p_1)(1-p_3)) \\
& & + (1-p_2)\min (p_1 p_3, 1 - p_1 p_3) = 0.1850, \\
P_e^{(2)}(\Pi_{2}(X_1, X_2, X_3)| X_3) & = & p_3 \min ((1-p_1)(1-p_2), 1 - (1-p_1)(1-p_2)) \\
& & + (1-p_3)\min (p_1 p_2, 1 - p_1 p_2) = 0.1632. 
\end{eqnarray*}

Thus, it appears that the structure of the optimal solution when we seek approximate computation given a fixed number of bits, is somewhat different from the optimal strategy for zero error computation.

\subsubsection{Special case of the parity function} 
While the structure of the optimal strategy for the approximate computation of threshold functions remains elusive, the case of the parity function is solvable. In this section, we show that an intuitive greedy strategy is optimal for the approximate computation of the parity function. To begin with, the parity function of $n$ Boolean variables labeled $X_1$ through $X_n$ is defined as follows:
\begin{displaymath}
\Phi(\mathbf{X}^n) := \left\{\begin{array}{ll}0 \quad \textrm{if } \sum_{i}X_i \textrm{ is even}\\ 1 \quad \textrm{if } \sum_{i}X_i \textrm{ is odd.}\end{array}\right.
\end{displaymath}

We consider the computation of $\Phi(\mathbf{X}^n)$ in a broadcast scenario where $X_i \sim Bern(p_i)$. If we seek exact computation, the problem becomes trivial since each node must transmit its bit. Hence, we will consider approximate computation of parity under the conditional entropy metric. Let us suppose that nodes are allowed to exchange upto $(n - \theta)$ bits. Let $H^{(\theta)}(\Phi(\mathbf{X}^n))$ denote the  minimum conditional entropy of the function after $(n-\theta)$ bits are exchanged. The dynamic programming equation analogous to (\ref{dyn_prog_entropy}) is 
\begin{eqnarray}
H^{(\theta)}(\Phi(\mathbf{X}^n)) & = & \min_{i}\left\{p_i H^{(\theta)}(\Phi(\mathbf{X}^n_{-i})) + (1-p_i)H^{(\theta)}(\Phi(\mathbf{X}^n_{-i})) \right\} \nonumber \\
& = & \min_{i}\left\{H^{(\theta)}(\Phi(\mathbf{X}^n_{-i}))\right\} \label{dyn_prog_entropy_parity}
\end{eqnarray}
with the boundary condition $H^{(\theta)}(\Phi(\mathbf{X}^{\theta})) = h (P(\Phi(\mathbf{X}^{\theta}) = 1))$. One can derive the solution to (\ref{dyn_prog_entropy_parity}) and hence deduce the optimal strategy for approximate computation of parity.
\begin{theorem}\label{thm_parity}
In order to minimize the conditional entropy of $\Phi(\mathbf{X}^n)$ after $(n-\theta)$ bits are exchanged, it is optimal for the node with highest binary entropy to transmit first. Subsequently, the node with the next highest entropy transmits, and so on until $(n- \theta)$ bits are transmitted.
\end{theorem}
\textbf{Proof: } First, we note that (\ref{dyn_prog_entropy}) implies that the optimal strategy for approximate computation of $\Phi(\mathbf{X}^n)$ is not data-dependent. Indeed, if node $i$ transmits first, irrespective of whether $X_i = 0$ or $X_i =1$, we have the residual task of computing $\Phi(\mathbf{X}^n_{-i})$ given at most $(n- \theta -1)$ bits. Thus, the optimal strategy can be specified \textit{a priori} and does not depend on the particular values of the bits exchanged. Further, if our objective is to minimize the conditional entropy after $(n - \theta)$ bits, we are only interested in determining the optimal subset of nodes that must transmit, and the order of transmission within this set is irrelevant. Thus, we have
\begin{displaymath}
H^{(\theta)}(\Phi(\mathbf{X}^n)) = \min_{\begin{array}{ll}S \subseteq \{1, \ldots, n\} \\ |S| = n- \theta \end{array} }  H(\Phi(\mathbf{X}^n)|\mathbf{X}_S).
\end{displaymath}

Let $A = \{a_1, a_2, \ldots ,a_{n-\theta}\}$ be an optimal set of nodes. We claim that $A$ consists of nodes with the $(n - \theta)$ highest entropies among the $n$ nodes. Suppose not. Then there exists nodes $a^{*} \notin A$ and $a_{i} \in A$ such that $H(p_{a^{*}}) > H(p_{a_i})$. Consider the set $A^{*} : = (A \setminus \{a_i\}) \bigcup \{a^{*}\}$. 
\begin{eqnarray*}
H(\Phi(\mathbf{X}^n)|\mathbf{X}_{A^{*}})& = & H(\Phi(\mathbf{X}^n)|\mathbf{X}_{A \setminus \{a_i\}}, X_{a^{*}}) \\
& = & H(\Phi(\mathbf{X}^n_{-((A \setminus \{a_i\}), a^{*})})) \\
& = & H(p_{a_i} P(\Phi(\mathbf{X}^n_{-(A, a^{*})})) = 1) + (1-p_{a_i}) P(\Phi(\mathbf{X}^n_{-(A, a^{*})})) = 0)) \\
& \leq & H(p_{a^{*}} P(\Phi(\mathbf{X}^n_{-(A, a^{*})})) = 1) + (1-p_{a^{*}}) P(\Phi(\mathbf{X}^n_{-(A, a^{*})})) = 0)) \\
& = & H(\Phi(\mathbf{X}^n_{-((A \setminus a^{*}), a_i)})),
\end{eqnarray*}
which contradicts the assumption that $A$ is an optimal subset.  Thus, under the toptimal strategy, the set of transmitting nodes must be those with the highest entropies. A candidate strategy which achieves this is one where nodes transmit in decreasing order of their binary entropies. $\Box$
\section{Concluding remarks}
We have addressed the problem of computing symmetric Boolean functions in a collocated wireless sensor network. In the worst case setting, we have derived optimal strategies for computing threshold functions, and order optimal strategies with optimal preconstant for interval functions. The approach presented can be extended in two directions. First, one can consider non-Boolean alphabets and functions which depend only on $\sum_{i} X_i$. Alternately, one can consider non-Boolean functions of a Boolean alphabet. The fooling set lower bound and the strategy for achievability can be generalized to both these cases. 

In the average case setting, we have considered some sequential decision problems, that arise in the context of optimal distributed computation of Boolean functions of random data. The broadcast nature of the medium forces nodes to communicate sequentially, and the challenge is to order nodes' transmissions so as to both exploit the structure of the function and the knowledge of the underlying distribution. 

For single instance computation of Boolean threshold functions, we have shown the surprising result that the optimal strategy has an elegant structure, which depends only on the ordering of the marginal probabilities, but not on their exact values. The extension to the case of block computation is harder. However, we have derived the optimal strategy for a restricted class of coherent strategies, which we conjecture to be optimal in general. The proof technique presented allows a unified treatment of these two problems, and also allows extension to alternate pulse models of communication where nodes transmit pulses of energy.

Finally, we have considered the problem of approximate function computation, where we are given a fixed number of bits and seek to minimize the error in the estimate of the function. We have shown that this problem is more complicated and the optimal strategy lacks the structure that we observed in the case of exact computation. However, for the special case of the parity function, a simple greedy strategy is optimal. There remain several open problems concerning optimal computation in wireless sensor networks, including the case of correlated measurements, and generalizing the sequential decision making approach to handle more general functions.

\appendix
\subsection{Proofs of Lemma \ref{thm_seq_gen}}	
\subsubsection{Proof of Lemma \ref{lemma_three}}\label{sec_proof_three}
First, let us suppose $k=0$. In this case
\begin{displaymath}
S^{(1)}_{n+1, 0,i}(\mathbf{X}^{n+1}) = (p_1 - p_i)C(\Pi_n(\mathbf{X}^{n+1}_{-(1,i)})) + (1-p_{1})C(\Pi_{n+1}(\mathbf{X}^{n+1}_{-1})) - (1-p_{i})C(\Pi_{n+1}(\mathbf{X}^{n+1}_{-i})) = 0
\end{displaymath}
However, by assumption, we have $0 \leq (1-p_1)f(p_i)- (1-p_i)f(p_1)$. 

Next, consider the case where $k \neq 0$.
\begin{eqnarray}
\hspace{-0.2in}& & (p_{k+1} - p_{i})C(\Pi_{n-k}(\mathbf{X}^{n+1}_{-(k+1, i)})) + (1-p_{k+1})C(\Pi_{n-k+1}(\mathbf{X}^{n+1}_{-(k+1)})) - (1-p_{i})C(\Pi_{n-k+1}(\mathbf{X}^{n+1}_{-i})) \nonumber \\
\hspace{-0.2in}& = & (p_{k+1} - p_i)\left[ f(p_k) + p_k C(\Pi_{n-k-1}(\mathbf{X}_{-(k, k+1, i)}^{n+1})) + (1-p_k) C(\Pi_{n-k}(\mathbf{X}^{n+1}_{-(k, k+1, i)}))\right]  \nonumber \\
\hspace{-0.2in}& &  + (1-p_{k+1})\left[f(p_k) + p_k C(\Pi_{n-k}(\mathbf{X}^{n+1}_{-(k, k+1)})) + (1- p_k)C(\Pi_{n-k+1}(\mathbf{X}^{n+1}_{-(k, k+1)}))\right] \nonumber \\
\hspace{-0.2in}& & -(1- p_i)\left[f(p_k) + p_k C(\Pi_{n-k}(\mathbf{X}^{n+1}_{-(k, i)})) + (1- p_k) C(\Pi_{n-k+1}(\mathbf{X}^{n+1}_{-(k, i)}))\right] \label{proof_three_a} \\
\hspace{-0.2in}& = & p_k \left[(p_{k+1} - p_i)C(\Pi_{n-k-1}(\mathbf{X}_{-(k, k+1, i)}^{n+1})) \right. \nonumber \\
\hspace{-0.2in}& & \hspace{1.5in} \left. + (1-p_{k+1}) C(\Pi_{n-k}(\mathbf{X}^{n+1}_{-(k, k+1)}))- (1-p_i) C(\Pi_{n-k}(\mathbf{X}^{n+1}_{-(k, i)}))\right] \nonumber \\
\hspace{-0.2in}& & +(1-p_k)\left[(p_{k+1} - p_i)C(\Pi_{n-k}(\mathbf{X}_{-(k, k+1, i)}^{n+1})) \right. \\ 
\hspace{-0.2in}& & \hspace{1.5in}\left. + (1-p_{k+1}) C(\Pi_{n-k+1}(\mathbf{X}^{n+1}_{-(k, k+1)})) - (1-p_i)C(\Pi_{n-k+1}(\mathbf{X}^{n+1}_{-(k, i)}))\right]  \nonumber \\
\hspace{-0.2in}& \leq & p_k \left[(p_{k+1} - p_i)C(\Pi_{n-k-1}(\mathbf{X}_{-(k, k+1, i)}^{n+1})) \right. \nonumber \\
\hspace{-0.2in}& & \hspace{0.3in} \left. + (1-p_{k+1}) C(\Pi_{n-k}(\mathbf{X}^{n+1}_{-(k, k+1)})) - (1-p_i)C(\Pi_{n-k}(\mathbf{X}^{n+1}_{-(k, i)}))\right] + (1-p_k)S^{(1)}_{n, k-1, i-1}(\mathbf{X}^{n+1}_{-k}) \nonumber \\
\hspace{-0.2in}& \leq & p_k \left[(p_{k+1} - p_i)C(\Pi_{n-k-1}(\mathbf{X}_{-(k, k+1, i)}^{n+1})) \right. \nonumber \\
\hspace{-0.2in}& & \hspace{1.5in} \left. + (1-p_{k+1}) C(\Pi_{n-k}(\mathbf{X}^{n+1}_{-(k, k+1)})) - (1-p_i) C(\Pi_{n-k}(\mathbf{X}^{n+1}_{-(k, i)}))\right] \nonumber \\
\hspace{-0.2in}& & + (1-p_k)\left[(1-p_{k+1})f(p_i) - (1-p_i)f(p_{k+1})\right] \label{proof_three_b} 
\end{eqnarray}
\begin{eqnarray}
\hspace{-0.2in}& = & p_k \left[(p_{k+1} - p_i) C(\Pi_{n-k-1}(\mathbf{X}^{n+1}_{-(k, k+1, i)})) + (1-p_{k+1}) C(\Pi_{n-k}(\mathbf{X}^{n+1}_{-(k, k+1)}))\right.  \nonumber \\
\hspace{-0.2in} & & \hspace{0.5in} \left. -(1-p_i)[f(p_{k+1} + p_{k+1} C(\Pi_{n-k-1}(\mathbf{X}^{n+1}_{-(k,k+1, i)})) + (1-p_{k+1}) C(\Pi_{n-k}(\mathbf{X}^{n+1}_{-(k, k+1, i)}))]\right] \nonumber \\
\hspace{-0.2in} & & + (1-p_k)\left[(1-p_{k+1})f(p_i) - (1-p_i)f(p_{k+1})\right]\label{proof_three_c} \\
\hspace{-0.2in}& = & p_k(1-p_{k+1})\left[ C(\Pi_{n-k}(\mathbf{X}^{n+1}_{-(k, k+1)})) - p_i C(\Pi_{n-k-1}(\mathbf{X}^{n+1}_{-(k, k+1, i)}))\right.  - \left.(1-p_i) C(\Pi_{n-k}(\mathbf{X}^{n+1}_{-(k, k+1, i)}))\right] \nonumber \\
\hspace{-0.2in}&  & -p_{k}(1-p_i)f(p_{k+1}) + (1-p_k)\left[(1-p_{k+1})f(p_i) - (1-p_i)f(p_{k+1})\right] \nonumber \\
\hspace{-0.2in}& \leq & p_k(1-p_{k+1})f(p_i) - (1-p_i)f(p_{k+1}) + (1-p_k)(1-p_{k+1})f(p_i) \label{proof_three_d} \\
\hspace{-0.2in}& = & (1-p_{k+1}) f(p_i) - (1-p_i) f(p_{k+1}) \nonumber
\end{eqnarray}
Equation (\ref{proof_three_a}) follows from the optimal ordering for computing $\Pi_{n-k}(\mathbf{X}^{n+1}_{-(k+1,i)})$, $\Pi_{n-k+1}(\mathbf{X}^{n+1}_{-(k+1)})$ and $\Pi_{n-k+1}(\mathbf{X}^{n+1}_{-i})$, which is true by the induction hypothesis for $m = n$. The inequality (\ref{proof_three_b}) follows from the induction hypothesis that $S^{(1)}_{n,k-1,i}(\mathbf{X}^{n+1}_{-k}) \leq (1-p_{k+1})f(p_i) - (1-p_i)f(p_{k+1})$. Equality in (\ref{proof_three_c}) and (\ref{proof_three_d}) follows from the optimal ordering for computing $\Pi_{n-k}(\mathbf{X}^{n+1}_{-(k, i)})$  and $\Pi_{n-k}(\mathbf{X}^{n+1}_{-(k, k+1)})$ respectively. $\Box$
\subsubsection{Proof of Lemma \ref{lemma_four}}\label{sec_proof_four}
First, let us suppose $k=n$. In this case
\begin{displaymath}
S^{(2)}_{n+1, n,i}(\mathbf{X}^{n+1}) = (p_i - p_{n+1})C(\Pi_0(\mathbf{X}^{n+1}_{-(i,n+1)})) + p_{n+1}C(\Pi_{0}(\mathbf{X}^{n+1}_{-(n+1)})) - p_i C(\Pi_{0}(\mathbf{X}^{n+1}_{-i})) = 0.
\end{displaymath}
However, by assumption, we have $0 \leq p_{n+1}f(p_i)- p_i f(p_{n+1})$. 

Next, consider the case where $k < n$.
\begin{eqnarray}
\hspace{-0.2in}& & (p_{i} - p_{k+1})C(\Pi_{n-k}(\mathbf{X}^{n+1}_{-(i, k+1)})) + p_{k+1}C(\Pi_{n-k}(\mathbf{X}^{n+1}_{-(k+1)})) - p_i C(\Pi_{n-k}(\mathbf{X}^{n+1}_{-i})) \nonumber \\
\hspace{-0.2in}& = & (p_{i} - p_{k+1})\left[ f(p_{k+2}) + p_{k+2} C(\Pi_{n-k-1}(\mathbf{X}_{-(i, k+1, k+2)}^{n+1})) \right. \nonumber + \left. (1-p_{k+2}) C(\Pi_{n-k}(\mathbf{X}^{n+1}_{-(i, k+1, k+2)}))\right] \nonumber \\
\hspace{-0.2in}& &+ p_{k+1}\left[f(p_{k+2}) + p_{k+2} C(\Pi_{n-k-1}(\mathbf{X}^{n+1}_{-(k+1, k+2)})) + (1- p_{k+2})C(\Pi_{n-k}(\mathbf{X}^{n+1}_{-(k+1, k+2)}))\right] \nonumber \\
\hspace{-0.2in}& &- p_i\left[f(p_{k+2}) + p_{k+2} C(\Pi_{n-k-1}(\mathbf{X}^{n+1}_{-(i, k+2)})) + (1- p_{k+2}) C(\Pi_{n-k}(\mathbf{X}^{n+1}_{-(i, k+2)}))\right] \label{proof_four_a} \\
\hspace{-0.2in}& = & p_{k+2} \left[(p_{i} - p_{k+1})C(\Pi_{n-k-1}(\mathbf{X}_{-(i, k+1, k+2)}^{n+1})) \right. \nonumber \\
\hspace{-0.2in}& & \hspace{1.9in} \left. + p_{k+1} C(\Pi_{n-k-1}(\mathbf{X}^{n+1}_{-(k+1, k+2)})) - p_i C(\Pi_{n-k-1}(\mathbf{X}^{n+1}_{-(i, k+2)}))\right] \nonumber \\
\hspace{-0.2in}& & +(1-p_{k+2})\left[(p_{i} - p_{k+1})C(\Pi_{n-k}(\mathbf{X}_{-(i, k+1, k+2)}^{n+1})) \right. + p_{k+1} C(\Pi_{n-k}(\mathbf{X}^{n+1}_{-(k+1, k+2)})) \nonumber \\
\hspace{-0.2in}& & \hspace{1.9in} \left.+ p_{k+1} C(\Pi_{n-k}(\mathbf{X}^{n+1}_{-(k+1, k+2)})) -  p_i C(\Pi_{n-k}(\mathbf{X}^{n+1}_{-(i, k+2)}))\right] \nonumber \\
\hspace{-0.2in}& \leq & (1-p_{k+2})\left[(p_{i} - p_{k+1})C(\Pi_{n-k}(\mathbf{X}_{-(i, k+1, k+2)}^{n+1}))\right. \nonumber \\
\hspace{-0.2in}& & \hspace{0.9in} + \left. p_{k+1} C(\Pi_{n-k}(\mathbf{X}^{n+1}_{-(k+1, k+2)})) - p_i C(\Pi_{n-k}(\mathbf{X}^{n+1}_{-(i, k+2)}))\right] + p_{k+2}\left[S^{(2)}_{n, k, i}(\mathbf{X}^{n+1}_{-(k+2)})\right] \nonumber \\
\hspace{-0.2in}& \leq & (1 - p_{k+2})\left[(p_{i} - p_{k+1})C(\Pi_{n-k}(\mathbf{X}_{-(i, k+1, k+2)}^{n+1})) \right. \nonumber \\
\hspace{-0.2in}& & \hspace{1.9in} \left. + p_{k+1} C(\Pi_{n-k}(\mathbf{X}^{n+1}_{-(k+1, k+2)})) -  p_i C(\Pi_{n-k}(\mathbf{X}^{n+1}_{-(i, k+2)}))\right] \nonumber \\
\hspace{-0.2in} & & + p_{k+2}\left[ p_{k+1} f(p_i) - p_i f(p_{k+1})\right] \label{proof_four_b} 
\end{eqnarray}
\begin{eqnarray}
\hspace{-0.2in}& = & (1- p_{k+2})\left[(p_{i} - p_{k+1})C(\Pi_{n-k}(\mathbf{X}_{-(i, k+1, k+2)}^{n+1})) + p_{k+1} C(\Pi_{n-k}(\mathbf{X}^{n+1}_{-(k+1, k+2)})) \right. \nonumber \\
\hspace{-0.2in} & & \hspace{0.5in}\left. - p_i[ f(p_{k+1}) + p_{k+1} C(\Pi_{n-k-1}(\mathbf{X}^{n+1}_{-(i, k+1, k+2)})) + (1-p_{k+1}) C(\Pi_{n-k}(\mathbf{X}^{n+1}_{-(i, k+1, k+2)}))]\right] \nonumber \\
\hspace{-0.2in} & & + p_{k+2}\left[ p_{k+1} f(p_i) - p_i f(p_{k+1})\right] \label{proof_four_c} \\
\hspace{-0.2in}& = & (1- p_{k+2})p_{k+1}\left[ C(\Pi_{n-k}(\mathbf{X}^{n+1}_{-(k+1, k+2)})) - p_i C(\Pi_{n-k-1}(\mathbf{X}^{n+1}_{-(i, k+1, k+2)}))\right. \nonumber \\
\hspace{-0.2in}& & \hspace{3.0in}- \left. (1-p_i) C(\Pi_{n-k}(\mathbf{X}^{n+1}_{-(i, k+1, k+2)}))\right]  \nonumber \\
\hspace{-0.2in}&  & - (1-p_{k+2})p_i f(p_{k+1}) + p_{k+2}\left[ p_{k+1} f(p_i) - p_i f(p_{k+1})\right] \nonumber \\
\hspace{-0.2in}& \leq & (1- p_{k+2})p_{k+1}f(p_{i})  - p_i f(p_{k+1}) + p_{k+2} p_{k+1} f(p_i)  \label{proof_four_d} \\
\hspace{-0.2in}& = & p_{k+1} f(p_i) - p_i f(p_{k+1}) \nonumber
\end{eqnarray}

Equation (\ref{proof_four_a}) follows from the optimal ordering for computing $\Pi_{n-k}(\mathbf{X}^{n+1}_{-(i, k+1)})$, $\Pi_{n-k}(\mathbf{X}^{n+1}_{-(k+1)})$ and $\Pi_{n-k}(\mathbf{X}^{n+1}_{-i})$, which follows from the induction hypothesis for $m = n$. The inequality (\ref{proof_four_b}) follows from the induction hypothesis that $S^{(2)}_{n,k,i}(\mathbf{X}^{n+1}_{-(k+2)}) \leq  p_{k+1} f(p_i) - p_i f(p_{k+1})$. Equations (\ref{proof_four_c}) and (\ref{proof_four_d}) follow from the optimal ordering for computing $\Pi_{n-k}(\mathbf{X}^{n+1}_{-(i, k+2)})$  and $\Pi_{n-k}(\mathbf{X}^{n+1}_{-(k+1, k+2)})$ respectively. $\Box$
\subsubsection{Proof of Lemma \ref{lemma_one}}\label{sec_proof_one}
First, we observe that 
\begin{multline}
T_{n+1, k, i}(\mathbf{X}^{n+1}) - S^{(1)}_{n+1, k, i}(\mathbf{X}^{n+1}) = p_{k+1} C(\Pi_{n-k}(\mathbf{X}^{n+1}_{-(k+1)})) - p_{i}C(\Pi_{n-k}(\mathbf{X}^{n+1}_{-i})) \\
- (p_{k+1} - p_{i})C(\Pi_{n-k}(\mathbf{X}^{n+1}_{-(k+1, i)})). \nonumber 
\end{multline}
Thus it is enough to show that 
\begin{multline}
p_{k+1}C(\Pi_{n-k}(\mathbf{X}^{n+1}_{-(k+1)})) - p_{i}C(\Pi_{n-k}(\mathbf{X}^{n+1}_{-i})) \\
\leq (p_{k+1} - p_{i})C(\Pi_{n-k}(\mathbf{X}^{n+1}_{-(k+1, i)}))  + p_{k+1} f(p_i) - p_i f(p_{k+1}) \qquad \textrm{ for } i \geq k+2. \nonumber
\end{multline}
First, observe that for $k =n$, the statement is vacuously true since $i \geq n+2$ is impossible. Hence, let us suppose that $k<n$. We have
\begin{eqnarray}
\hspace{-0.5in}& & p_{k+1}C(\Pi_{n-k}(\mathbf{X}^{n+1}_{-(k+1)})) - p_{i}C(\Pi_{n-k}(\mathbf{X}^{n+1}_{-i}))  \nonumber \\
\hspace{-0.5in}& = &p_{k+1}\left[ f(p_{k+2}) + p_{k+2}C(\Pi_{n-k-1}(\mathbf{X}^{n+1}_{-(k+1, k+2)})) + (1-p_{k+2})C(\Pi_{n-k}(\mathbf{X}^{n+1}_{-(k+1,k+2)}))\right] \nonumber \\
\hspace{-0.5in}& &  -p_i \left[f(p_{k+1} + p_{k+1}C(\Pi_{n-k-1}(\mathbf{X}^{n+1}_{-(k+1, i)})) + (1-p_{k+1})C(\Pi_{n-k}(\mathbf{X}^{n+1}_{-(k+1, i)}))\right] \label{proof_one_a} \\
\hspace{-0.5in}& = & p_{k+1}\left[f(p_{k+2}) + p_{k+2}C(\Pi_{n-k-1}(\mathbf{X}^{n+1}_{-(k+1,k+2)})) - p_i C(\Pi_{n-k-1}(\mathbf{X}^{n+1}_{-(k+1, i)}))\right] \nonumber \\
\hspace{-0.5in}& & +  p_{k+1}(1 - p_{k+2})C(\Pi_{n-k}(\mathbf{X}^{n+1}_{-(k+1,k+2)})) - p_i(1-p_{k+1})C(\Pi_{n-k}(\mathbf{X}^{n+1}_{-(k+1, i)})) - p_i f(p_{k+1}) \nonumber \\
\hspace{-0.5in}&  \leq & p_{k+1}\left[f(p_i) + (1-p_i)C(\Pi_{n-k}(\mathbf{X}^{n+1}_{-(k+1, i)})) - (1-p_{k+2})C(\Pi_{n-k}(\mathbf{X}^{n+1}_{-(k+1,k+2)})) \right] \nonumber \\
\hspace{-0.5in}& & + p_{k+1}(1 - p_{k+2})C(\Pi_{n-k}(\mathbf{X}^{n+1}_{-(k+1,k+2)})) - p_i(1-p_{k+1})C(\Pi_{n-k}(\mathbf{X}^{n+1}_{-(k+1, i)})) - p_i f(p_{k+1})\label{proof_one_b} \\
\hspace{-0.5in}&=& (p_{k+1} - p_i)C(\Pi_{n-k}(\mathbf{X}^{n+1}_{-(k+1, i)}))  + p_{k+1} f(p_i) - p_i f(p_{k+1}) \nonumber
\end{eqnarray}
Equation \ref{proof_one_a} follows from the optimal order for computing $\Pi_{n-k}(\mathbf{X}^{n+1}_{-(k+1)})$ and $\Pi_{n-k}(\mathbf{X}^{n+1}_{-i})$. The inequality in \ref{proof_one_b} follows from the induction hypothesis $T_{n, k, i}(\mathbf{X}^{n+1}_{-(k+1)}) \leq f(p_i) - f(p_{k+2})$. $\Box$
\subsubsection{Proof of Lemma \ref{lemma_two}}\label{sec_proof_two}
First, we observe that 
\begin{multline}
T_{n+1, k, i}(\mathbf{X}^{n+1}) - S^{(2)}_{n+1, k, i}(\mathbf{X}^{n+1}) = (1-p_{k+1})C(\Pi_{n-k+1}(\mathbf{X}^{n+1}_{-(k+1)})) - (1-p_{i})C(\Pi_{n-k+1}(\mathbf{X}^{n+1}_{-i})) \\
- (p_{i} - p_{k+1})C(\Pi_{n-k}(\mathbf{X}^{n+1}_{-(i,k+1)})). \nonumber 
\end{multline}
Thus it is enough to show that 
\begin{multline}
(1-p_{k+1})C(\Pi_{n-k+1}(\mathbf{X}^{n+1}_{-(k+1)})) - (1-p_{i})C(\Pi_{n-k+1}(\mathbf{X}^{n+1}_{-i})) \\
\leq (p_{i} - p_{k+1})C(\Pi_{n-k}(\mathbf{X}^{n+1}_{-(i,k+1)}))  + (1-p_{k+1}) f(p_i) - (1-p_i) f(p_{k+1}) \qquad \textrm{ for } i \leq k. \nonumber
\end{multline}
First, observe that for $k =0$, the statement is vacuously true since $i \leq 0$ is impossible. Hence, let us suppose that $k>0$. We have
\begin{eqnarray}
& &  (1-p_{k+1})C(\Pi_{n-k+1}(\mathbf{X}^{n+1}_{-(k+1)})) - (1-p_{i})C(\Pi_{n-k+1}(\mathbf{X}^{n+1}_{-i})) \nonumber \\
&  = &  (1-p_{k+1})\left[f(p_k) + p_{k}C(\Pi_{n-k}(\mathbf{X}^{n+1}_{-(k, k+1)})) + (1-p_{k})C(\Pi_{n-k+1}(\mathbf{X}^{n+1}_{-(k,k+1)}))\right] \nonumber \\
& &  -(1- p_i) \left[f(p_{k+1}) + p_{k+1}C(\Pi_{n-k}(\mathbf{X}^{n+1}_{-(i, k+1)})) + (1-p_{k+1})C(\Pi_{n-k+1}(\mathbf{X}^{n+1}_{-(i,k+1)}))\right] \label{proof_two_a} \\
& = & (1- p_{k+1})\left[f(p_k) + (1- p_{k})C(\Pi_{n-k+1}(\mathbf{X}^{n+1}_{-(k,k+1)})) - (1-p_i) C(\Pi_{n-k+1}(\mathbf{X}^{n+1}_{-(i, k+1)}))\right] \nonumber \\
& &  + p_{k}(1 - p_{k+1})C(\Pi_{n-k}(\mathbf{X}^{n+1}_{-(k,k+1)})) - p_{k+1}(1-p_{i})C(\Pi_{n-k}(\mathbf{X}^{n+1}_{-(i,k+1)})) - (1-p_i) f(p_{k+1})\nonumber \\
& \leq & (1-p_{k+1})\left[ f(p_i) + p_i C(\Pi_{n-k}(\mathbf{X}^{n+1}_{-(i,k+1)})) - p_k C(\Pi_{n-k}(\mathbf{X}^{n+1}_{-(k,k+1)}))\right] \nonumber \\
& & + p_{k}(1 - p_{k+1})C(\Pi_{n-k}(\mathbf{X}^{n+1}_{-(k,k+1)})) - p_{k+1}(1-p_{i})C(\Pi_{n-k}(\mathbf{X}^{n+1}_{-(i,k+1)})) \label{proof_two_b} \\
&  =  & (p_{i} - p_{k+1})C(\Pi_{n-k}(\mathbf{X}^{n+1}_{-(i,k+1)})) + (1-p_{k+1})f(p_i) - (1-p_i)f(p_{k+1}) \nonumber
\end{eqnarray}
Equation (\ref{proof_two_a}) follows from the optimal order for computing $\Pi_{n-k+1}(\mathbf{X}^{n+1}_{-(k+1)})$ and $\Pi_{n-k+1}(\mathbf{X}^{n+1}_{-i})$. The inequality in (\ref{proof_two_b}) follows from the induction hypothesis $T_{n, k-1, i}(\mathbf{X}^{n+1}_{-(k+1)}) \leq f(p_i) - f(p_k)$ $\Box$.
\bibliographystyle{unsrt}
\bibliography{defense_biblio}
\end{document}